\documentclass[sigconf,10pt]{acmart} 
\newcommand{\figref}[1]{Figure~\ref{#1}}
\newcommand{\secref}[1]{Section~\ref{#1}}

\newcommand{\heading}[1]{\smallskip\noindent{\bf #1}}

\usepackage{multirow}
\usepackage{epstopdf}

\usepackage{url} 

\usepackage[utf8]{inputenc}

\usepackage{graphicx}
\usepackage{times}

\usepackage[english]{babel}
\usepackage{blindtext}
\usepackage{endnotes}
\usepackage{color,hyphenat}
\usepackage{listings}
\usepackage{booktabs}
\usepackage{algorithm,algpseudocode}
\usepackage[labelformat=simple]{subcaption}
\usepackage{stfloats}

\copyrightyear{2023}
\acmYear{2023}
\acmConference[ACM MobiCom '23]{The 29th Annual International Conference on Mobile Computing and Networking}{October 2--6, 2023}{Madrid, Spain}
\acmBooktitle{The 29th Annual International Conference on Mobile Computing and Networking (ACM MobiCom '23), October 2--6, 2023, Madrid, Spain}\acmDOI{10.1145/3570361.3592520}
\acmISBN{978-1-4503-9990-6/23/10}

\renewcommand\footnotetextcopyrightpermission[1]{} 
\setcopyright{none}
\definecolor{dkgreen}{rgb}{0,0.6,0}
\definecolor{gray}{rgb}{0.5,0.5,0.5}
\definecolor{mauve}{rgb}{0.58,0,0.82}

\algnewcommand\And{\textbf{and }}
\algnewcommand\Or{\textbf{or }}

\newcommand{\name}{\textit{\textbf{Scrolls}}}
\newcommand{\stitle}{\name{}: Rolling Flexible Surfaces for Wideband Wireless}

\newcommand{\edit}[1]{\textcolor{black}{#1}}
\newcommand{\myedit}[1]{\textcolor{black}{#1}}

\begin{document}

\title[\stitle]{Softly, Deftly, Scrolls Unfurl Their Splendor: \\ Rolling Flexible Surfaces for Wideband Wireless}

\author{Ruichun Ma}
\affiliation{\institution{Yale University}}

\author{R. Ivan Zelaya}
\affiliation{\institution{Yale University}}

\author{Wenjun Hu}
\affiliation{\institution{Yale University}}


\begin{abstract}

With new frequency bands opening up, emerging wireless IoT devices are capitalizing on an increasingly divergent range of frequencies. 
However, existing coverage provisioning \edit{practice} is often tied to specific standards and frequencies. There is little \textit{shareable} wireless infrastructure for concurrent links on different frequencies, across networks and standards.

This paper presents \name, a 
frequency-tunable \textit{soft} smart surface system to enhance wideband, multi-network coverage. \name' hardware comprises many rows of rollable thin \edit{plastic film}, 
each attached with flexible copper strips. 
When rolled to different lengths, the copper strips act as wire antennas reflecting signals on the corresponding frequencies. The surface control algorithm determines the \edit{unrolled strip} lengths for link enhancement by probing the search space efficiently. We build a set of distributed, composable \name\ prototypes and deploy them in an office. 
Extensive evaluation shows that \name\ can adapt the antenna lengths effectively to provide link enhancement across diverse standards on sub-6~GHz bands. For concurrent links on 900~MHz (LoRa), 2.4~GHz (Wi-Fi), 3.7~GHz, and 5~GHz, \name\ can provide received signal strength gains to all links simultaneously, by a median of 4~dB and up to 10~dB.




\end{abstract}



\maketitle
\section{Introduction}
\label{s:intro}
\begin{figure}[t]
   \centering
  \begin{subfigure}[b]{0.53\columnwidth}
     \includegraphics[width=0.97\columnwidth]{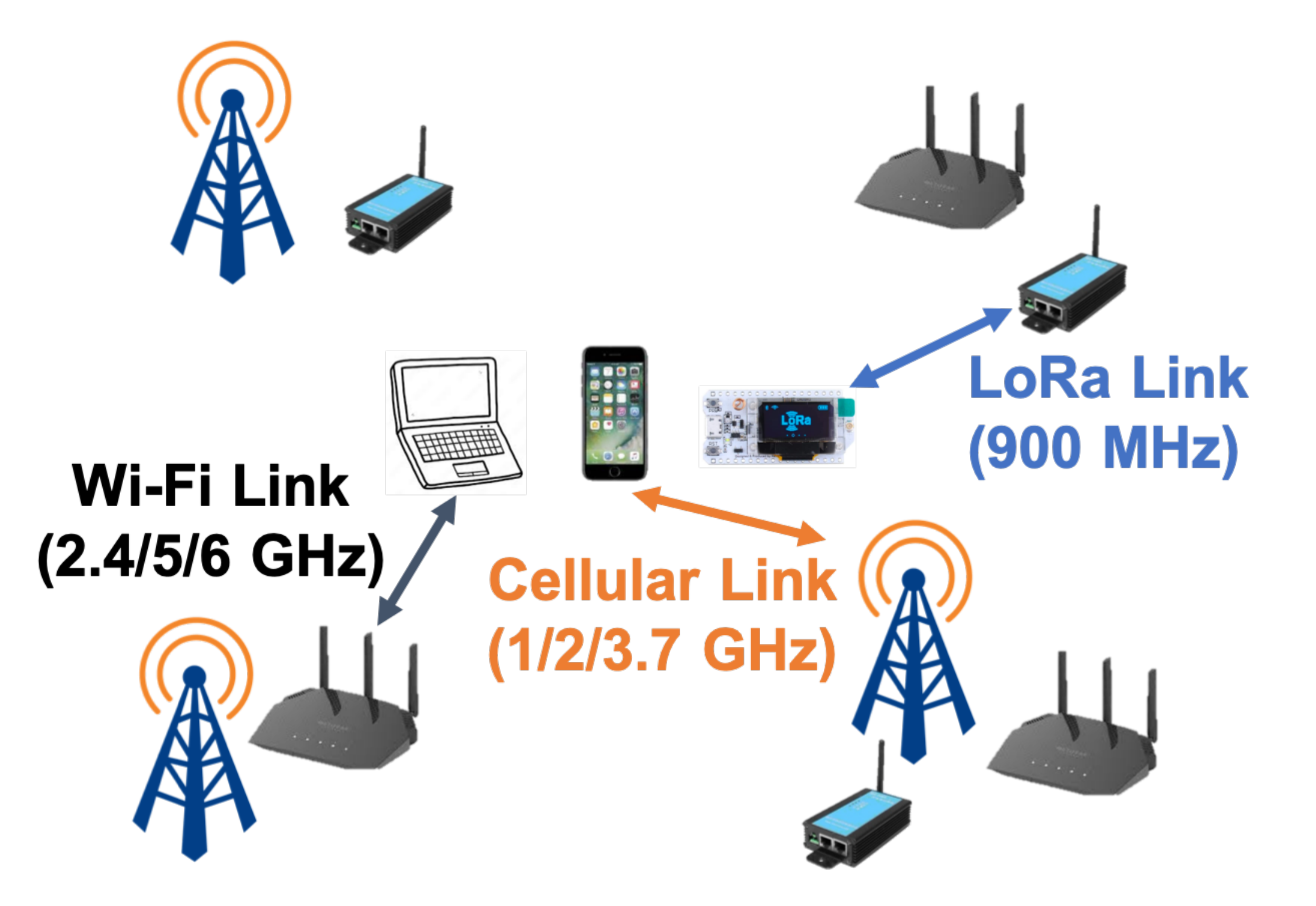}
     \caption{}
     \label{fig:intro-scenario-existing}
  \end{subfigure}
  \begin{subfigure}[b]{0.44\columnwidth}
     \includegraphics[width=0.97\columnwidth]{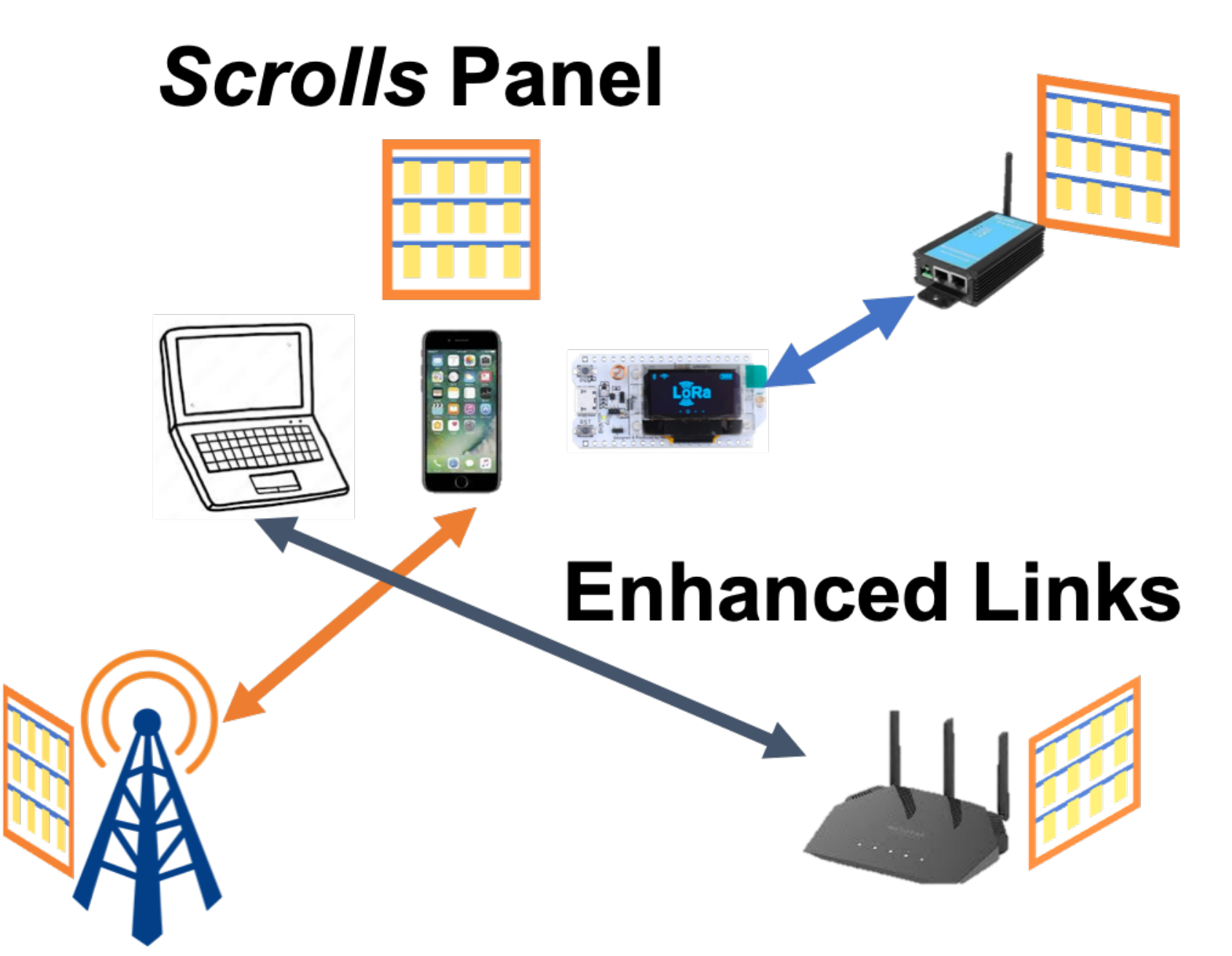}
     \caption{}
     \label{fig:intro-scenario-scrolls}
  \end{subfigure}
  \caption{\textbf{(a)} Provisioning wireless coverage currently requires densely deployed standard- and frequency-specific infrastructure. \textbf{(b)} \name{} provides simultaneous coverage enhancement for diverse standards and frequencies, thus reducing hardware deployment cost and complexity.}
  \label{fig:intro-scenario}
\end{figure}

The number of wireless Internet of Things (IoT) devices has been growing exponentially for years. \myedit{Many IoT scenarios today are mostly coverage-bound, i.e., the devices are limited by unfavorable channel conditions, but do not always require high throughput or high mobility.}
\edit{These} devices often have limited resources due to their small form factors and low-cost hardware, necessitating infrastructure support, e.g., for massive connectivity. 
However, it is not just the sheer number and the per-device resource constraints that pose scalability challenges. Alongside that growth is increasing diversification of these devices across the board, in terms of the target applications, device form factor and capability, and the frequency bands they reside on (\secref{sec:motivation}). 

Currently, pervasive wireless coverage is achieved via both sophisticated designs and dense deployment of the ``access nodes'' to counter signal degradation, which can be costly and complex even for \textit{a single standard and frequency}. \edit{For example, cellular base stations for 5G, (multi-band, backward compatible) access points for Wi-Fi, and LoRa gateways for LoRa, have to be deployed separately, often connected to separate access networks but may cover the same geographic regions.}
To make matters worse, improving coverage for multiple co-located technologies nevertheless requires one separate set of additional hardware per standard (\figref{fig:intro-scenario-existing}), despite achieving similar purposes of link enhancement. 


Fundamentally, there are few existing solutions to provide standard-agnostic, multi-frequency coverage enhancement that is shareable among devices, across networks for different standards. Developing such solutions is non-trivial. Standard-agnosticism requires signal-level operations, below and independent from the physical layer symbol definition and framing. Multi-frequency support requires the hardware to adapt to a wide range of resonant frequencies, i.e., generating tangible frequency responses over a large range.

Hence an emerging direction is to deploy \textit{smart surface} (like) systems~\cite{press-hotnets17,laia-nsdi19,rfocus, scattermimo, llama,lava-sigcomm21}. These systems actuate passing wireless signals to control the end-to-end signal propagation behavior and improve the perceived channel conditions at the receiver endpoints. They can be viewed as more general-purpose physical layer relays. 
Since smart surfaces operate at the signal level, these are the closest we have to achieving standard-agnostic coverage provisioning. 

However, existing surface designs cannot support 
wide-ranging frequencies, because they are implicitly tuned to a small frequency range. They follow a similar high-level design, i.e., comprising a large array of \textit{surface elements}. Each element includes one or two antennas (or metallic patterns acting as antennas) connected to programmable RF circuit components for signal manipulation. The frequency range supported is limited by both the resonant frequency of the antennas and the operating ranges of the circuit components.

Are there simple refinements to achieve wideband, multi-network support? 
It may appear that the existing surface designs could be revised by using wideband/multi-band components or assembling multiple narrowband designs. However, both suffer from several limitations and require an impractically large number of hardware elements to support concurrent links on multiple frequencies (\figref{fig:motiv-sim-tunability}). 
We discuss further in \secref{sec:motiv-tunable} and experimentally test a wideband antenna-based surface in \secref{sec:eval:freq-tunable} (\figref{fig:eval-freq-tunability}).
Instead, we propose \name{}, a frequency-tunable reflective smart surface system that enhances wideband, multi-network coverage. Crucially, its surface elements can be individually tuned to all frequencies continuously over a wide range, hence the system as a whole can more easily tailor to multiple links transmitting simultaneously on different frequencies.
This way, \name{} directly augments the propagation environment and minimizes the use of expensive standard and frequency-specific active devices, e.g., densely deployed base stations, APs, gateways, or repeaters (\figref{fig:intro-scenario-scrolls}).
\myedit{This is complementary to adding standard-specific active devices. 
The goal is to have low-cost, long-lasting shareable infrastructure support.}

At its core, \name\ is a collection of flexible wire antennas that can be rolled to suitable lengths to match different operating frequencies (\secref{sec:design}). We use flexible antennas and no RF circuit components to minimize the frequency range constraints. Programmability is achieved via mechanical rolling actuation instead. 
Our design currently targets the sub-6~GHz bands, since these are widely used for 5G/6G IoT scenarios due to low-cost RF hardware and longer signal propagation ranges. We build prototypes to mimic a distributed surface deployment; this supports a large coverage area with relatively small surface sizes and facilitates diverse experiment setups (\secref{sec:impl}).

Extensive evaluation (\secref{sec:eval}) shows \name\ is effective, especially for concurrent link support. 
\name\ can improve the received signal strength for individual 915~MHz LoRa, 2.4~GHz Wi-Fi, 3.7~GHz and 5~GHz links. 
For example, \name\ achieves 47\% TCP throughput increase on average and up to 8$\times$ for Wi-Fi links. 
For concurrent links on 915~MHz, 2.4~GHz, 3.7~GHz and 5~GHz, \name\ can provide a median gain of 4~dB and up to 10~dB gain in the received signal strength across multiple groups of concurrent links. 
We also analyze the inner workings of \name{}, its responsiveness and performance stability.

To summarize, this paper makes the following contributions. First, we highlight the need to design a wideband, multi-network coverage enhancement solution, in light of increasing adoption of diverse frequency bands. Second, we propose \name, a new \textit{soft} smart surface design with a tunable frequency response over sub-6~GHz bands and the associated control algorithms. Third, extensive evaluation over a composable prototype spotlights the agility of such a surface design, especially when catering to multiple concurrent links on different frequencies, across standards and networks.

\section{Background and Motivation}
\label{sec:motivation}


Until around 10 to 15 years ago, the most frequently discussed radio bands in commodity usage were the unlicensed ISM bands used by Wi-Fi (first the 2.4~GHz band, then the 5~GHz band) and the licensed cellular bands (various bands from sub-GHz to 2~GHz). Since then, more frequency bands have been opened up as the initial users upgrade to new technologies. For example, TV whitespace bands~\cite{tv-white-space} (700~MHz in the US) became available after TV broadcast moved to digital. 900~MHz is now unlicensed in the US, used by LoRa, Z-Wave, and 802.11ah. The 802.11ax standard initially targeted the same usual Wi-Fi bands, but is now being extended to include 6~GHz, as Wi-Fi 6E. Cellular technologies are eyeing even more diverse ranges, from mmWave ranges (e.g., 24~GHz, 60~GHz) already being piloted for 5G deployment to (sub-)~terahertz ranges (300~GHz to THz) being actively studied in various research and engineering efforts. In the lower frequency spectrum, the C-band~\cite{5g-c-band} (3.7 to 4~GHz) has been featured prominently in 
news articles~\cite{what-is-c-band,c-band-att-verizon}.
There is also plan to free up more spectrum in the future~\cite{freeSpectrum-Biden-2023}.

Although mmWave and higher bands provide sufficient spectrum resources for high-bitrate communication, sub-6~GHz is still crucial for resource constrained wireless (IoT) devices due to the long communication range and low-cost hardware. 
As a result, we are seeing an increasingly divergent set of wireless devices and corresponding operating frequencies in the sub-6~GHz range. These low-cost devices with limited hardware capability often suffer from difficult channel conditions, hence unstable connectivity or reduced coverage. 
Thus, it is more pressing and challenging than ever to provide infrastructure support for wireless connectivity in view of diversifying 
standards and frequencies.


\subsection{Existing coverage enhancement solutions}

\heading{Standard-specific infrastructure. }
The de facto coverage provisioning strategy is to (densely) deploy multiple sets of infrastructure per standard and frequency, e.g., cellular base stations, Wi-Fi APs, and standard-specific repeaters, in the hope of these covering all spots reliably. Multi-band devices rely on multi-band \textit{antennas} if only one band is used at a time, or additionally multiple \textit{radios} (e.g., dual-band Wi-Fi APs) to use those bands concurrently, increasing the hardware cost to multiples of single-band radios. 
%
%
While this has worked so far, it is increasingly unscalable as more standards and frequency bands are employed at the same time and over the same geographic area, e.g., Wi-Fi, LoRa, Z-Wave, ZigBee, Thread, LTE, 5G .... To extend coverage or improve link quality for various wireless devices, we need to install more base stations for cellular, more access points for Wi-Fi, more gateways for LoRa, ... To make the matter worse, the hardware needs to be upgraded regularly to new versions.
Although the hardware cost can be amortized over a large network of endpoints, 
in practice individual IoT networks are not always large enough for this amortization effect.
\edit{Ideally}, we need a shareable coverage enhancement solution, which can serve different standards on wide ranging frequencies.


\heading{Standard-agnostic smart surfaces.} 
A recent proposal is to deploy \textit{smart surfaces} (or surface-like systems) as part of 6G infrastructure~\cite{next-g-alliance-report,empower-tech-roadmap,6g-vision-hexa} to enhance wireless coverage. Early end-to-end prototypes~\cite{laia-nsdi19, rfocus, scattermimo, lava-sigcomm21, llama, RFlens, cho2022mmwall} actuate passing wireless signals via signal-level operations, e.g., phase shifts, on-off reflection amplitude control, polarization rotation, or signal amplification. 
Although only LAVA~\cite{lava-sigcomm21} explicitly demonstrates such standard-agnostic features in an end-to-end fashion, smart surfaces in general operate at the signal level and aim to improve the perceived channel conditions at the wireless endpoints without targeting a specific wireless standard. \edit{They cannot yet obviate the need for standard-specific \textit{endpoint} devices, but they can minimize the use of such standard-specific endpoints by enhancing link quality or range}. 

However, the extent of standard-agnosticism is limited to the operating frequency band the surface is designed for, due to the constraints from the antennas and RF components used. Phase-shifter-based systems are more constrained, as the amount of phase shift possible is dictated tightly by the operating frequency. Systems mainly augmenting the incident signal amplitude (such as RFocus~\cite{rfocus} and LAVA~\cite{lava-sigcomm21}) are more amenable to wideband operations, provided all system components work on a wide frequency range.


\subsection{Towards Frequency-Tunable Surfaces} 
\label{sec:motiv-tunable}
A natural question is whether we can extend existing surface designs \edit{to work on more frequencies}. Unfortunately, this is not trivial. Instead, we need to redesign the surface for \textit{frequency tunability}, i.e., to dynamically configure its operating frequency. 
In this section, we argue that wideband frequency tunability is essential \edit{for smart surfaces} to support concurrent links on wide-ranging frequencies. We discuss the limitations of potential wideband designs and compare them with a frequency tunable design in simulations.

\begin{figure*}[t]
   \centering
   \begin{subfigure}[b]{0.33\textwidth}
     \includegraphics[width=0.9\columnwidth]{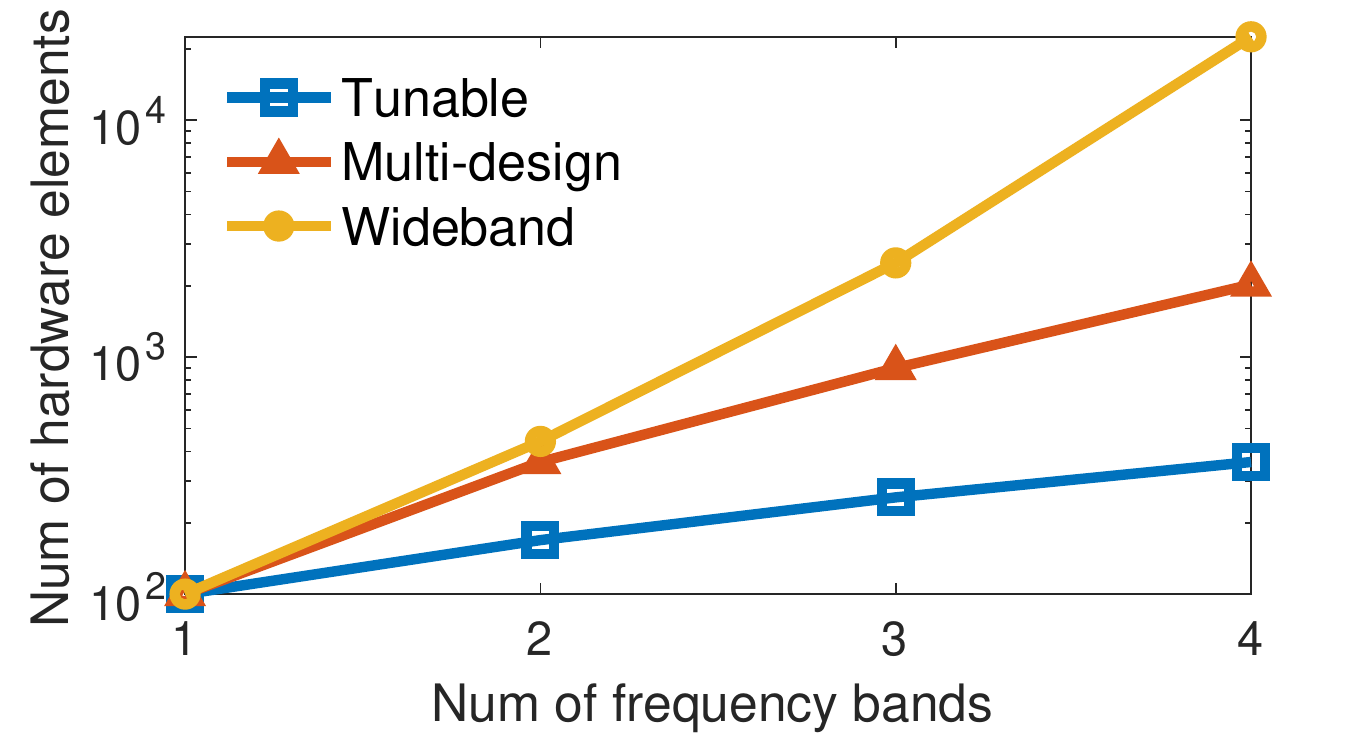}
     \caption{}
     \label{fig:motiv-sim-hardware-num}
  \end{subfigure}\hspace{1em}%
  \begin{subfigure}[b]{0.3\textwidth}
     \includegraphics[width=0.95\columnwidth]{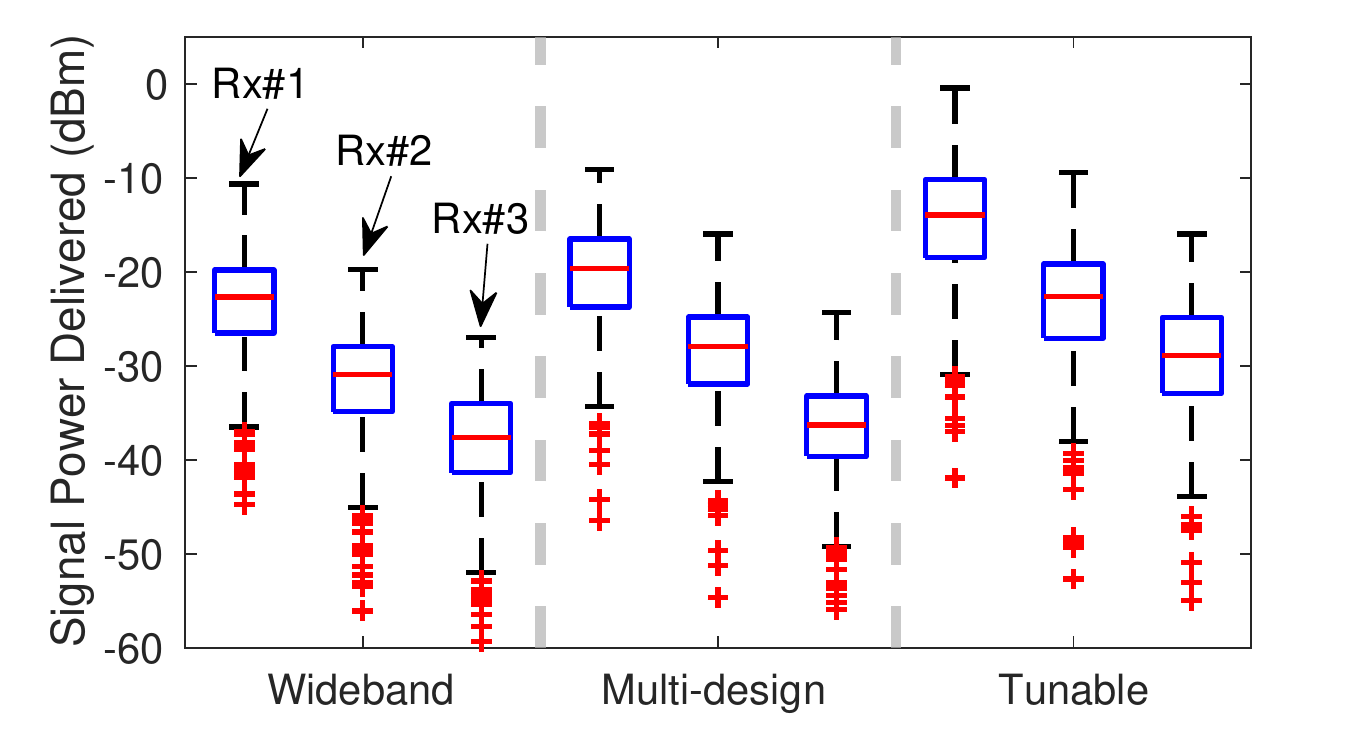}
     \caption{}
     \label{fig:motiv-sim-rss-boxplot}
  \end{subfigure}\hspace{1em}%
  \begin{subfigure}[b]{0.33\textwidth}
     \includegraphics[width=0.9\columnwidth]{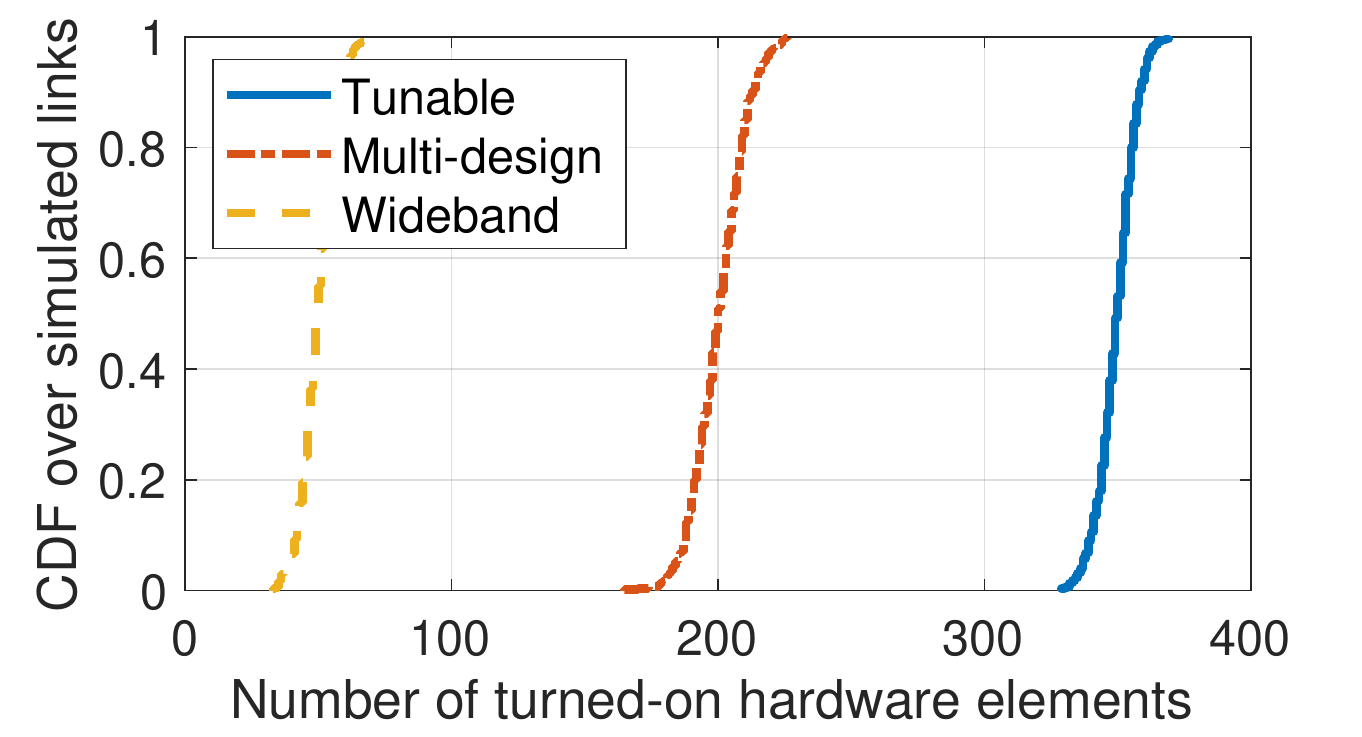}
     \caption{}
     \label{fig:motiv-sim-turn-on-num}
  \end{subfigure}
  \caption{Simulations showing the advantages of frequency tunability. \textmd{\textbf{(a)} The numbers of smart surface elements needed to achieve comparable performance when supporting multiple frequencies using different designs. When supporting receivers on 3 frequencies with 20$\times$20 surface element arrays, we compare \textbf{(b)} the signal power delivered from the surfaces and \textbf{(c)} the numbers of turned-on surface elements.} }
  \label{fig:motiv-sim-tunability}
\end{figure*}

\heading{Wideband/Multi-band elements.}
We may consider extending existing surface designs with wideband/multi-band components and antennas. Unfortunately, there are several issues with this approach. First, the hardware cost for wideband antennas and RF components is significantly higher compared to their single band counterparts given their design and fabrication complexity. Second, wideband/multi-band antennas often require a much larger size~\cite{antenna-theory-design, multiband-antennas}, to maintain good performance over the whole frequency range. This then necessitates a large spacing between adjacent antennas, leading to a sparser array of smart surface elements. A sparse array means lower beamforming efficiency and fewer deployable elements for a limited deployment area, both of which degrade the performance. The last but critical issue is that wideband/multi-band components would not permit frequency-selective controllability. We can only impose effects on all frequencies alike, so they exhibit an all-or-nothing behavior. Such frequency-selective control is nevertheless essential when supporting multiple concurrent links on different frequencies, as we show later with simulations (\figref{fig:motiv-sim-tunability}).


\heading{\edit{Multi-design integration.}} 
Another potential approach is to assemble \edit{and integrate} multiple element designs, such as \edit{geometrically} scaling the same design 
\edit{corresponding} to different frequencies and placing \edit{the frequency-specific variants} side by side. This way, each design caters to a specific frequency band. Commodity dual-band Wi-Fi routers typically adopt this approach. This way, we have frequency-selective controllability for a fixed set of frequencies at the expense of more hardware components. 

Still, there are several critical drawbacks. 
First, we can hardly predict the set of frequencies 
used at a given time or location. Suppose we 
target the three most widely used ISM bands, i.e., 900~MHz, 2.4~GHz, and 5~GHz. Most of the time, endpoints nearby may only use one frequency band, which would mean a huge waste of surface hardware. On the other hand, if endpoints use any other band, like cellular bands or 6~GHz, the design would fail to support them. Second, it is hard to determine the right way to assemble multiple designs. 
Placing multiple designs side by side could lead to sparse arrays for all frequencies or unwanted coupling effects between designs.
Lastly, 
we do not have the ability to dynamically allocate hardware resources to different frequencies, which eventually leads to suboptimal performance.


\heading{Simulation study.}
We run Matlab simulations 
to \edit{illustrate} 
the aforementioned shortcomings of wideband antenna and multi-design approaches. We \edit{view the signal propagation behavior from a surface-centric perspective and} simulate the behavior of three smart surface element arrays: (i) a wideband surface, where the elements can reflect signals on all bands alike; (ii) a multi-design surface, where we have multiple sets of elements, each set reflecting signals on a fixed frequency; and (iii) a frequency-tunable surface, where each element can choose to work on any band. We double the spacing between wideband surface elements as a moderate estimate for the size of wideband antenna. We do not \edit{simulate endpoint transmitters explicitly}, but randomly generate a direct path channel for the surface control algorithm to align \edit{the surface configuration} with. All surface antenna elements are considered isotropic with on-off control and we use the control algorithm in RFocus~\cite{rfocus}, i.e., turning on the element when its reflected signals' phases align with the direct link phase at the receiver; It is proved to be near-optimal when the number of elements is large. Finally, we measure the power delivered from the surface to the receiver endpoint.




In \figref{fig:motiv-sim-hardware-num}, we compare the number of surface elements needed to achieve the same received signal strength when supporting multiple links on different frequencies, using a 10$\times$10 \myedit{array supporting one link as the baseline}. The frequency tunable surface needs significantly less hardware deployed compared to the other designs, and its advantage becomes more pronounced with more frequencies considered. 
\edit{Smart surface deployment typically requires proximity to endpoints for optimal performance, but there is limited physical space for deployment.}
Thus, a massive surface deployment is not only costly, but also impractical in most cases. Note that we assume the multi-design surface has prior knowledge of what frequencies are used in each simulation, which is \edit{impractical}. In a practical deployment, we would need to pre-allocate far more multi-design elements than the simulation \edit{setup} to account for the worst case. 

Next, we compare the signal power \edit{delivered} by each design with the same element count (\figref{fig:motiv-sim-rss-boxplot}). The tunable array provides around 6~dB higher median power than the multi-design surface for all links, and 9~dB than the wideband antenna surface, when supporting 3 links on different frequencies with 20$\times$20 arrays. To understand the difference, \figref{fig:motiv-sim-turn-on-num} further plots the hardware utilization. With the ability to work on any frequency, the tunable surface can utilize most of its hardware efficiently.




\section{\name{} Design}
\label{sec:design}

We design \name\ to improve the channel conditions with an array of reflective copper strips as surface elements, which also have tunable operating frequency on sub-6~GHz bands. \name\ achieves a wide frequency tuning range at minimal hardware complexity by rolling the copper strips to harness the benefit of physical structural change. 

\subsection{Hardware}
\label{sec:design-hardware}
\name\ comprises a hierarchically organized set of copper strip-based reflective \textit{elements} on thin flexible sheets. Each sheet is rolled around a non-metallic rod (referred to as a \textit{roll}); rotating the rod extends or shrinks the exposed lengths of the copper strips, which tunes them to the corresponding resonant frequencies. Several rolls are assembled and fitted in a wooden frame, collectively referred to as a \textit{panel}.

\heading{Basic element design.} 
The operating frequency of an antenna is mainly determined by its physical dimension. Changing its dimension therefore naturally adjusts its operating frequency. To induce this dimension change easily, we use a flexible copper strip as the basic element, functioning as a half-wave dipole antenna. Its length should be around half of \edit{the wavelength for} the resonant frequency~\cite{antenna-theory-design}. Then we can adjust its operating frequency dynamically by changing its effective length with mechanical rolling actuation. The rolling actuation shields the rolled part of the copper strip, so the exposed length of each strip resonates with the desirable frequency. This way, our element design provides wideband frequency tunability at minimal hardware complexity.

\begin{figure*}[t]
    \centering
    \begin{subfigure}[b]{0.33\textwidth}
        \centering
        \includegraphics[width=0.99\columnwidth]{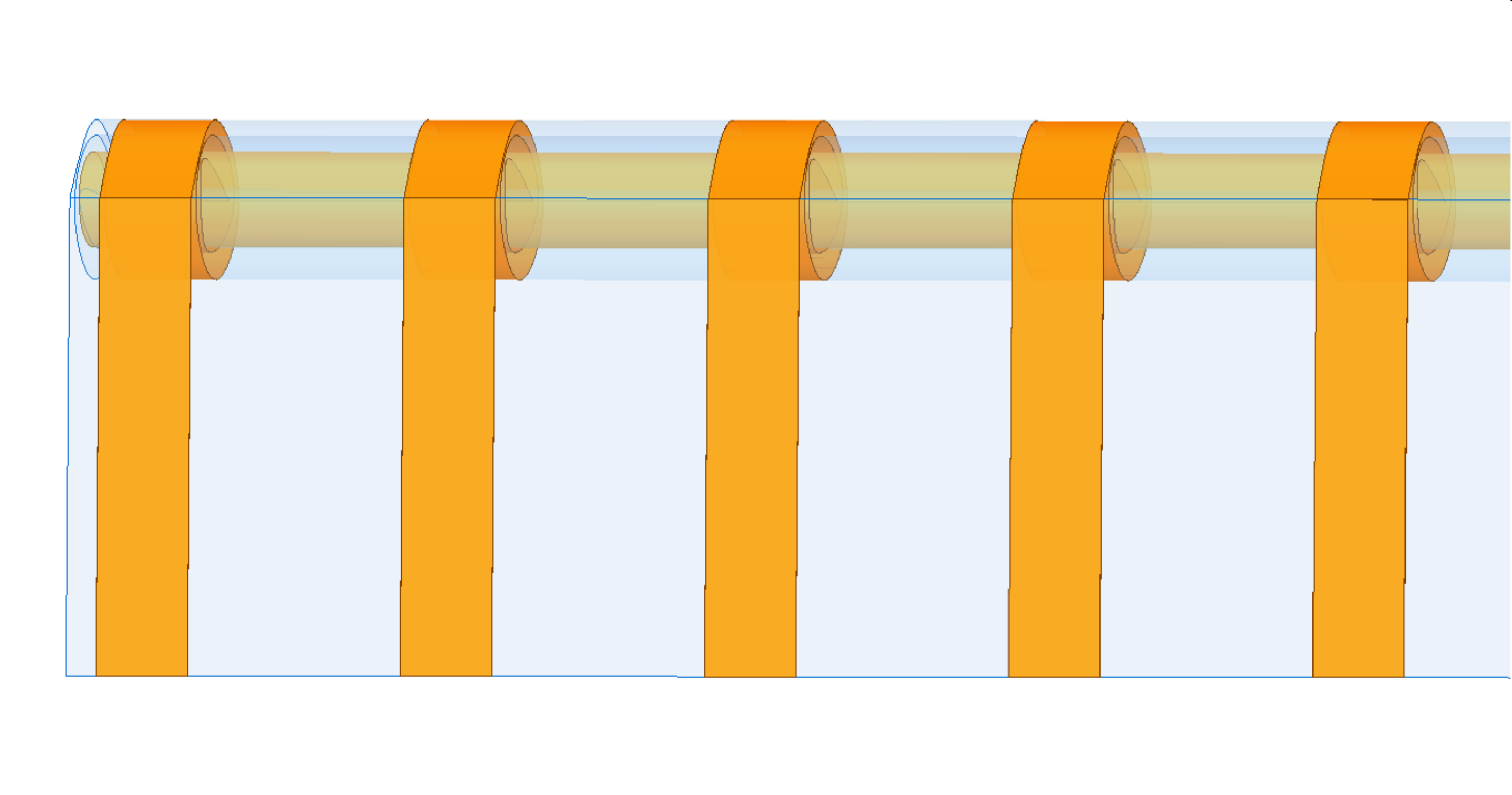}
        \caption{Rollable flexible antenna design}
        \label{fig:design-rolling-surf}
   \end{subfigure}\hspace{1em}%
   \begin{subfigure}[b]{0.3\textwidth}
        \centering
        \includegraphics[width=1\columnwidth]{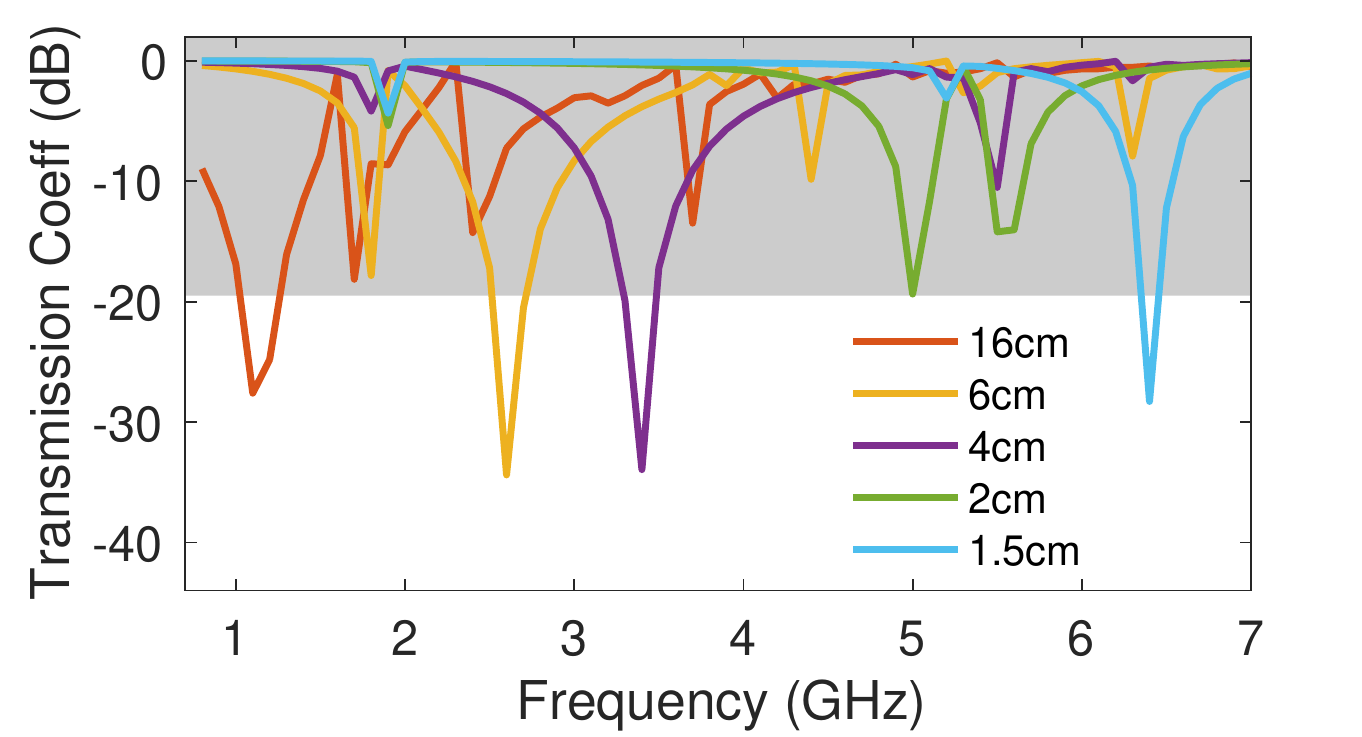}
        \caption{$S_{21}$ of rollable antenna design}
        \label{fig:hfss-S21-roll}
   \end{subfigure}\hspace{1em}%
   \begin{subfigure}[b]{0.33\textwidth}
        \centering
        \includegraphics[width=0.92\columnwidth]{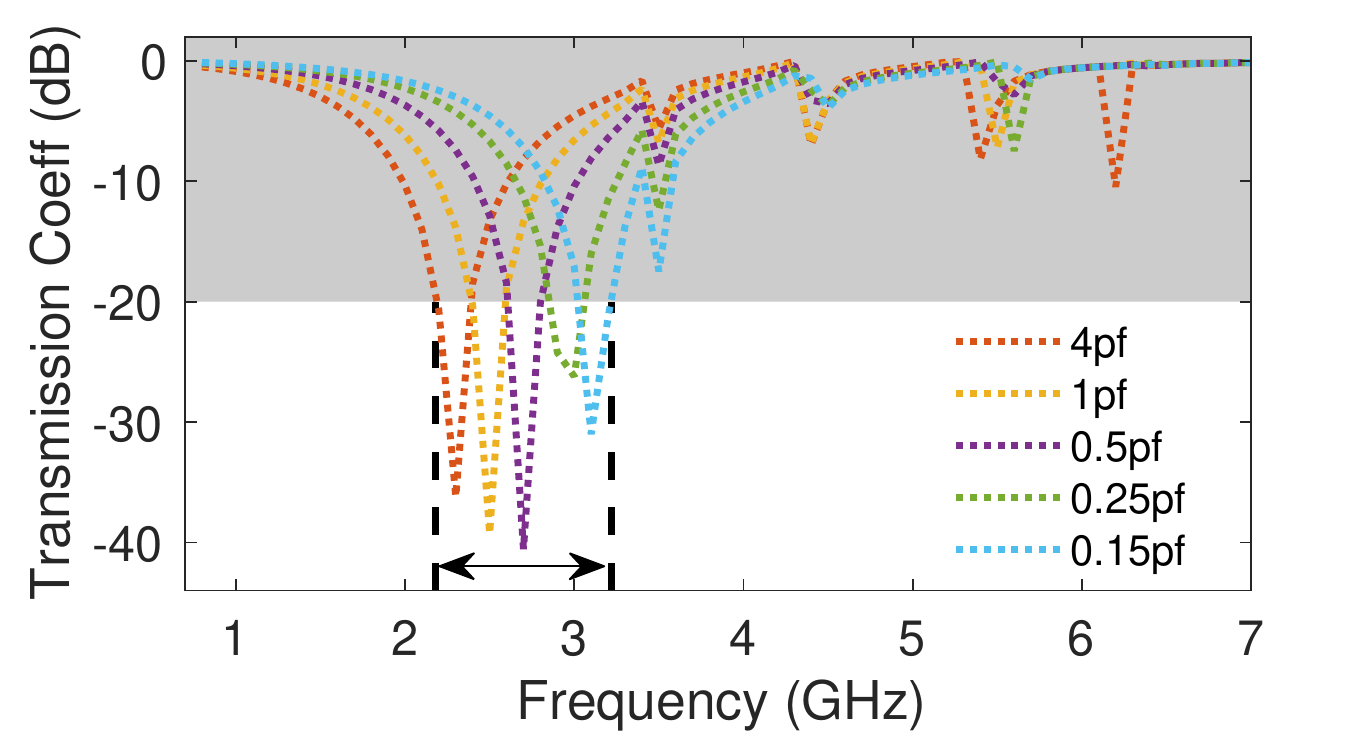}
        \caption{$S_{21}$ of varactor-based design}
        \label{fig:hfss-S21-varactor}
   \end{subfigure}
   \caption{\textmd{HFSS simulations of the \name\ design, compared to a varactor-based design. The lowest troughs are seen at the resonant frequencies. Setting -20~dB as the maximum transmission coefficient ($S_{21}$) to support signal reflection (white region of (b) and (c)), \name\ 
   shows a much wider tunable resonant frequency range from sub-GHz to over 6~GHz.}}
 \end{figure*}

\heading{Design simulation.}
We study the behavior of the above design via HFSS~\cite{hfss}, simulating an array of rolled antennas (\figref{fig:design-rolling-surf}) with varying exposed lengths for the copper strips. The transmission coefficient shown in \figref{fig:hfss-S21-roll} indicates how much signal power passes through the array, while the rest of the energy gets reflected.
\myedit{Our surface itself incurs negligible power loss, so any reduction in the transmission coefficient corresponds to an increase in the reflected signal strength.} 
We set the transmission coefficient target to -20~dB; below this threshold, the vast majority of the energy (99\% at -20~dB) is already reflected, and the benefit of achieving lower transmission values is negligible. 

By iteratively simulating various configurations, we identify suitable design parameters as follows: The copper strip \textit{width} is set to 6~mm for a suitable antenna operating bandwidth; The \textit{spacing between adjacent copper strips} is set to 3~cm to minimize potential element coupling effects; The completely unrolled length is set to 16~cm to cover 900~MHz to over 6~GHz. The simulations verify that 
\edit{adjusting exposed roll lengths changes the resonant frequency correspondingly, consistent with when these copper strips are viewed as half-wave antennas,}
despite some variance caused by the rolled parts of the strips. 
\myedit{However, we can not simulate real-world phase changes because HFSS can not simulate the multipath fading behavior from the elements' physical movement.} 
When the element length is 1~cm, there is no discernible effect over the frequency range of interest. Therefore, when an antenna is tightly rolled, we can consider it turned off (no reflection). In \secref{sec:eval:freq-tunable} (\figref{fig:eval-freq-tunability}), we experimentally verify our design and compare it to wideband antennas.

Our design supports any resonant frequency in the target range \textit{continuously}, not just a set of non-contiguous frequency bands. This provides forward compatibility for any band adopted in the future and provides some robustness to wear and tear, given our control algorithm can explore and find the length that works best (\secref{sec:design-control}). 


\heading{On-off amplitude control.} 
We mainly consider on-off amplitude control for each frequency since phase shifts brings limited extra gain, as shown in RFocus~\cite{rfocus}. Further, adding a phase shifter per antenna incurs significant complexity in both the hardware cost and the control algorithm. For instance, phase-shifting requires fine-grained channel state information (including phase values) and can potentially have a massive search space. Additionally, it is more sensitive to slight channel fluctuations. Nevertheless, the physical movement of flexible elements also changes the phase of reflected signals, providing another inherent degree of control.

\heading{Tunable frequency range limit.} 
Our design is intended for frequencies from 900~MHz to over 6~GHz. We currently do not target higher frequency bands (especially mmWave or terahertz), \edit{as they} require \edit{more aggressive} antenna beamforming. 
A frequency below 900~MHz can potentially be supported by increasing the maximal length of copper strips.

\heading{Comparison with a varactor-based design.}
A potential alternative to mechanical actuation is attaching a varactor diode to each antenna and applying a biasing voltage~\cite{reconfig-ant-deisgn, reconfig-ant-misc}. Varying the voltage can change the capacitance of the diode and shift the operating frequency. However, the tunable range using varactors is much smaller compared to changing the physical dimension of the antenna. We simulate a varactor-based frequency-tunable, fixed length antenna design with HFSS. \figref{fig:hfss-S21-varactor} shows only 1~GHz of tunable frequency range, which cannot satisfy our goal of covering most sub-6~GHz frequency bands. The ratio of the tunable range to the center frequency is around 40\% (1~GHz/2.5~GHz), which also agrees with the analysis in related work~\cite{frequency-agile-microstrip,mias2005varactorFSS}, while our system's ratio is around 200\% (6~GHz/3~GHz). Note that commercial varactors~\cite{varactor-spec} normally have a smaller capacitance range than what is used in the simulation. 
We also experimented with a smart surface prototype with varactors and its tunable frequency range was indeed very limited.
\myedit{As we aim to optimize for wideband support, we choose mechanical rolling actuation over electronic actuation for a much larger tunable frequency range.}

\heading{Array assembly and element spacing.} 
There are many ways to assemble elements into a large array. We want to strike a balance between sufficient control granularity and complexity; hence, we attach multiple strips onto one thin plastic sheet/surface and roll the entire \textit{\textbf{roll}} (\figref{fig:design-rolling-surf}). This grouping significantly reduces the hardware complexity at the expense of only having roll-wise control. We choose a moderate 40~cm roll length and a fixed, 3~cm element spacing according to our simulations, which leads to 14 elements per roll. We need a minimum spacing to avoid undesirable coupling effects, but want to upperbound the spacing to improve element beam efficiency. To simplify hardware design and control, we ensure the former but not optimize for the latter. Note that this spacing choice can be harder for other designs, such as wideband antenna based design, due to their large element size. A potential refinement is to incorporate research work on stretchable electronics~\cite{highly-stretchable-circuits}, which could vary both the element size and spacing simultaneously.


Instead of building one large, monolithic smart surface, we break the surface into several \textit{\textbf{panels}}, each including several separate rolls and an ESP32 micro-controller. The design choice to use panels enables both scalability and flexibility for deployment. More panels can be added to the system incrementally to improve the system coverage or performance. Meanwhile, since each panel functions independently, we can deploy panels in a distributed way, integrating them with furniture pieces or building structures to ensure proximity between the surface and the endpoints, which is vital for the performance of passive surfaces. 

 
\begin{figure}[t]
    \centering
    \includegraphics[width=0.8\columnwidth]{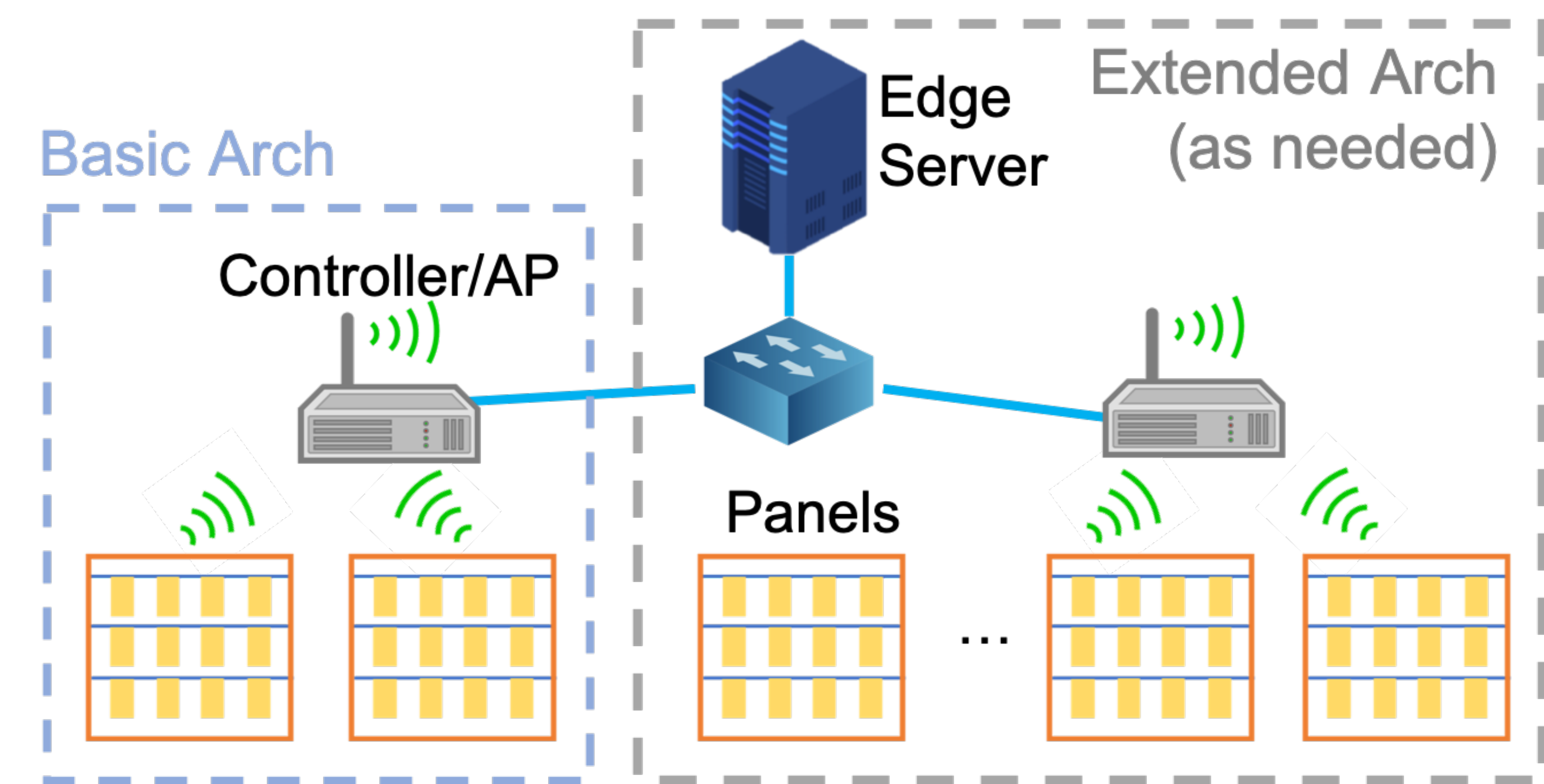}
    \caption{\name{} Control Architecture. \textmd{\name{} uses either a basic architecture, where a controller/AP controls the nearby panel(s), or an extended, tiered architecture for scalability.}}
    \label{fig:design-control-arch}
 \end{figure}

\heading{Control architecture. } 
\name{} panels can receive commands wirelessly from a close-by controller/AP to set the roll lengths. 
This works as the basic control architecture (\figref{fig:design-control-arch}). For a larger scale deployment, \name{} can adopt a tiered control architecture for scalability. Multiple controllers can control multiple groups of panels and forward commands from a server via Ethernet. The server runs the control algorithm based on the feedback from endpoints. Both the controllers and the server can be implemented as software modules running on existing infrastructure, for example, on Wi-Fi access points and an edge computing server. 
Consequently, the rolling panels would be the only extra hardware needed.


\subsection{Control Plane Action}
\label{sec:design-control}
We next present the control algorithm for \name{}. 
The goal is to set an appropriate length/state for each roll to maximize the received signal strength at endpoints.

\heading{Feedback acquisition.} 
\name{} requires guidance to direct the length change of each roll. Our current design does not include sensing capability to detect active transmissions or infer channel conditions, so we require explicit channel feedback from the endpoints, as is typical of previous passive surface designs~\cite{laia-nsdi19, scattermimo, rfocus, RFlens}. Since we aim to provide standard-agnostic support for resource constrained devices, the feedback we need from the receivers should be minimal and accessible across devices. Therefore, our algorithm requires only RSSI (Received Signal Strength Indicator) feedback, an aggregate metric over the channel that is likely to be more stable than finer-grained channel state information typically involving per-subcarrier signal strengths. From past work on Wi-Fi rate adaptation~\cite{effectivesnr-sigcomm10}, we know that RSSI more accurately captures the received signal quality for narrowband or spread spectrum modulation but not wideband OFDM or MIMO as used in Wi-Fi. However, the former is common for many IoT radios. We let endpoints send feedback proactively, using device-specific network interfaces. 

\heading{Basic algorithm.}  
There are several challenges for the control algorithm. First, the exact best roll length for a certain operating frequency is unknown. With limited fabrication precision, we cannot precisely map a given frequency to a specific roll length. Further, the phase of the reflected signal changes due to physical rolling movement. Thus, the algorithm needs to explore several lengths to find the best. We set a rough length range for each frequency band, for example, 1.5--4~cm for 5~GHz, 2--5~cm for 3.7~GHz, 5--9~cm for 2.4~GHz, and 10--16~cm for 900~MHz. We also set the appropriate granularity for each band to delineate different states: 0.5~cm for 5~GHz and 3.7~GHz, 1~cm for 2.4~GHz and 915~MHz. Such exploration leads to another challenge: the search space is too large to be explored exhaustively in a brute force manner. For $R$ rolls, each with $N$ states, the algorithm would need to explore $N^R$ configurations. We solve this problem by a greedy search, exploring one roll at a time, i.e., $N\times R$ configurations total. Lastly, to minimize the rolling distance, the algorithm sweeps over the potential lengths instead of switching between different lengths multiple times. To sum up, the basic algorithm consists of enumerating all rolls, one by one, by sweeping over their potential roll length range, and then measuring the effects (RSSI). The algorithm subsequently chooses the state for each roll that provides the best RSSI.

\heading{Speedup with group sweeping.} 
To further speed up the process, we employ a group sweeping algorithm, following the idea of group testing.
\myedit{For each test, we compare the RSSI measurements when the roll is sweeping over potential states/lengths to the RSSI when the roll is turned off. 
Our observation is that, despite the large search space, most roll states do not provide a measurable gain at a receiver due to phase (mis-)alignment or the beam direction.
Thus, the algorithm can test multiple rolls together to speed up.}
If no RSSI gain is observed at the receiver, all rolls tested are considered as non-helpful and turned off. This speeds up the search process. A challenge with this procedure is that signals reflected from multiple rolls may add destructively at the receiver, causing a false negative test result. Our solution is to use prior knowledge of panel placement. Given that each roll comprises several antennas, the reflected beam is directional. Hence, rolls from different panels are highly likely to reflect signals towards different directions such that perfect destructive interference at the endpoint receiver is improbable.


\algtext*{EndWhile}
\algtext*{EndIf}
\begin{algorithm}[t]
    \caption{The group sweeping algorithm}\label{alg:group-sweep}
    \begin{algorithmic}
    \While{not all rolls are tested}
        \State Randomly sample one untested roll from each panel
        \State Form a roll group G for test
        \State $rssi\_base \gets measure\_rssi() $
        \While{not reaching the end}
            \State Set rolls in G to the next length state
            \State $rssi\_current \gets measure\_rssi() $
            \If{$rssi\_current > rssi\_base$}
                \State Test rolls in G one by one until detect gain
                \State Leave untested rolls for next sweep
                \State \textbf{break}
            \EndIf
        \EndWhile
        \If{no gain was detected}
        \State  Mark all rolls in G as tested and set to off
        \EndIf
    \EndWhile
    \end{algorithmic}
\end{algorithm}    

\textbf{Algo}~\ref{alg:group-sweep} describes the algorithm in detail. First, we sample one roll from each panel to form a group and sweep over their potential length range. Before sweeping, we need to measure one RSSI value as the baseline. During the sweep, if we observe no RSSI gain, we mark all rolls as off and roll them to 1~cm. If we observe an RSSI gain, we run enumerative testing for the rolls in the current group until we find the roll that provides gain. We set the tested roll to the state that provides the highest gain. We leave the remaining untested rolls for the next group sweep. This is repeated until all rolls are set to appropriate states. To combat the inherent variance of RSSI feedback, we take the median across multiple measurements.

\heading{Configuration cache.}
\label{sec:cache-scheme}
We record the extended rolls and their lengths for a link and save these in a configuration cache. Instead of running the algorithm each time a link becomes active, we reuse the cached configuration first. As tested in \figref{fig:eval-single-link-gains}, loading the cached configurations provides similar enhancement to the ideal configuration. \name{} can check the validity of a cached configuration by resetting the relevant roll states, either periodically or proactively with a scenario-specific policy. The control algorithm is activated again when the cached configuration is no longer valid.

\heading{Handling concurrent links. } 
The above algorithm can be extended to cater to multiple concurrent links on the same or different frequencies, with two refinements. First, during the group sweep, we run enumerative testing when detecting gain from any link. Second, during the enumerating stage, we do not select any roll length state that causes performance loss at any link. We choose the length that provides the highest sum of gain from all links. \name{} may need to further consider fairness among all links, which depends on the specific application scenario. We leave a more intelligent algorithm that considers application scenarios for future exploration.

\heading{Blind spot connections.}
\myedit{Wireless links like Wi-Fi often remain connected over a long distance by using the lowest bitrate, but the throughput is too limited to be usable. In such cases, \name{} can still reliably acquire feedback and achieve link improvement. For new connections from blind spots, we can incorporate a periodic exploration process by varying the surface states to discover new connections.}

\heading{Mobility support.}
Our current system does not explicitly support mobility due to limited actuation speed. However, few IoT devices need to deal with high mobility and therefore any Doppler effect is negligible---i.e., the target frequencies would not change frequently. More likely, what we need to deal with is channel fluctuations from occasional movements (of either surrounding objects or the endpoints to another location). We show in \secref{sec:eval_micro} that \name\ is robust to this type of channel fluctuations (\figref{fig:eval-stability}). 

\myedit{Moreover, there is a trade-off between control complexity and adaptation speed to channel conditions. \name\ adapts to the operating frequency and channel fluctuations simultaneously with low-complexity mechanical actuation. If we were to support high mobility, we could additionally incorporate previous electronic control techniques~\cite{laia-nsdi19,rfocus,llama} to further adapt to channel variations, separately from tuning the operating frequency; the flexible surfaces can be fabricated as flexible PCBs and varactors can be added to the lower (unrolled) part of surfaces. However, this is beyond the scope of this paper.}




\section{Implementation}
\label{sec:impl}
 \begin{figure}[t]
    \centering
  \begin{subfigure}{0.49\columnwidth}
    \centering
    \includegraphics[scale=0.18]{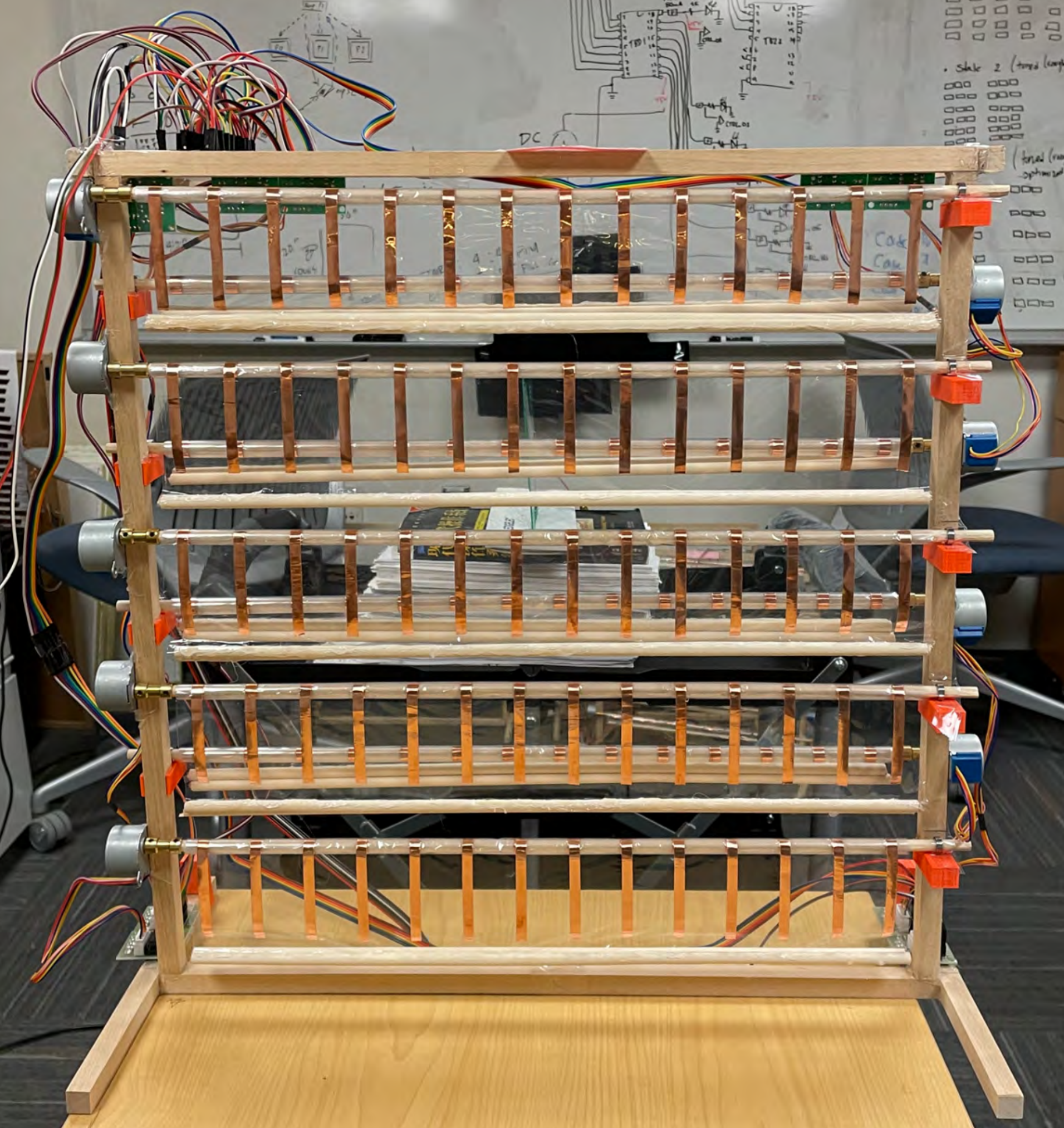}
    \caption{\name\ panel.}
    \label{fig:impl-panel-img}
  \end{subfigure}%
  \begin{subfigure}{0.49\columnwidth}
    \centering
    \includegraphics[scale=0.17]{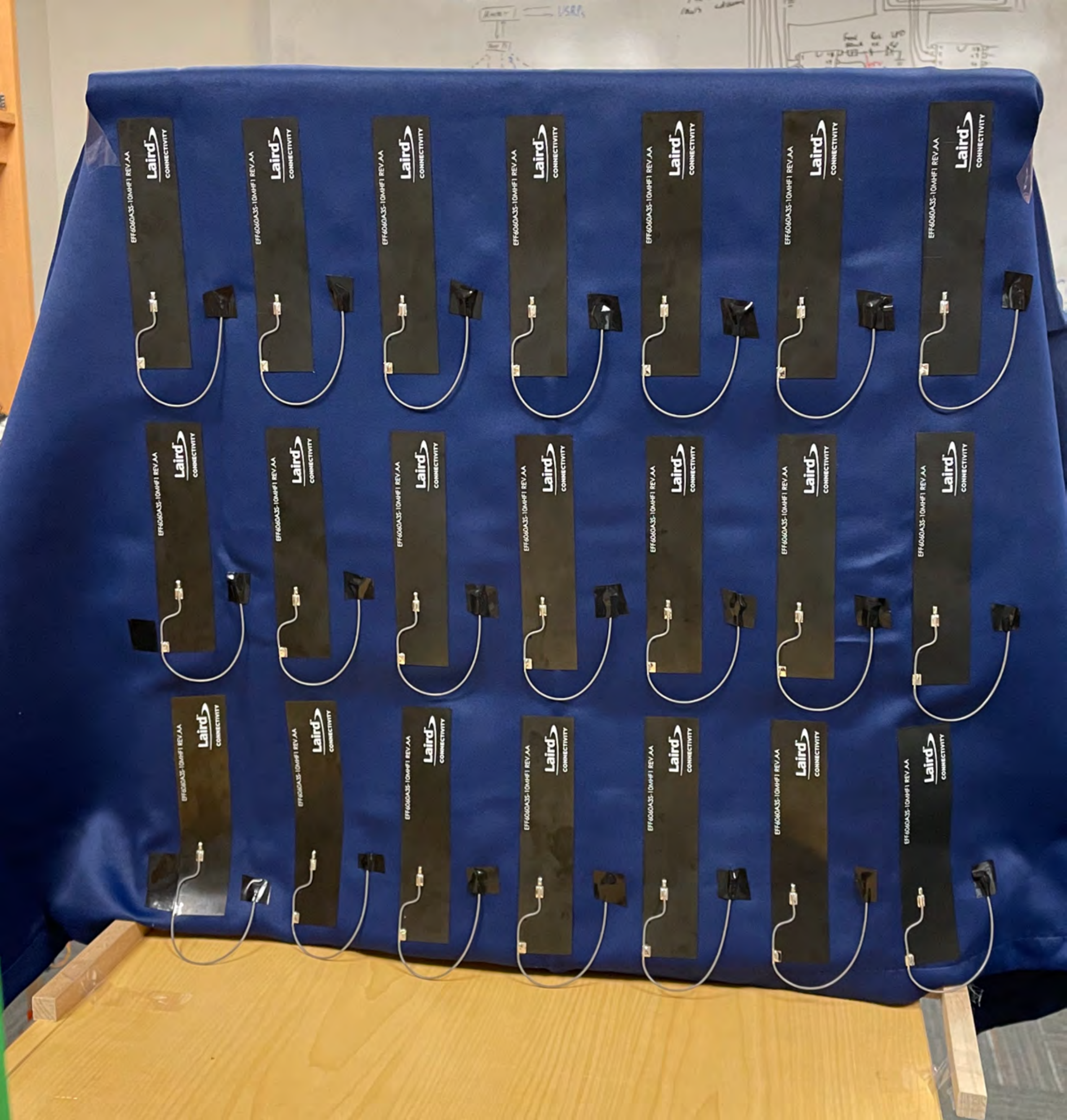}
    \caption{Wideband antenna surface.}
    \label{fig:eval-wideband-surf}
  \end{subfigure}
  \caption{Surface prototypes built for experiments.} 
\end{figure}

\heading{Panel material and fabrication.} 
\figref{fig:impl-panel-img} shows one panel with rolls and surfaces partly extended. We choose a 45~cm~$\times$~45~cm size for ease of deployment and testing. We fabricated the panels mostly with wood materials, including a square frame and rods with a 3~mm radius. One end of each rotating rod connects to a stepper motor~\cite{stepper_motor}, while the other end connects to a ball bearing. Each wooden rod hosts one roll of flexible plastic sheet with 1~mil (0.001~inch) thickness. Copper tapes, each 16~cm long and 6~mm wide, are attached to each sheet vertically, with 3~cm spacing. Each sheet is kept straight by two additional wooden rods, serving as weight, at the bottom of each sheet. 


Each panel hosts 9 rolls due to fabrication constraints. A denser placement may improve the efficiency of the hardware. When the rolls are sufficiently extended, any overlap between adjacent rolls could cause coupling. Therefore, all rolls are tightly rolled when idle to minimize possible coupling. When overlaying rolls is inevitable, the control algorithm inherently tests the effect of coupling and exploits it when favorable. Further, instead of calibrating each element precisely, we simply let the control algorithm compensate for potential errors by adjusting the lengths of each roll. 

\heading{Vertical surface placement.} 
Each panel currently uses vertically placed copper strips and rolls, so it 
may only work for vertically polarized signals. However, it is unlikely that signals from the endpoints will be perfectly mismatched with a \name\ panel, and the panel can still reflect the vertical component of \edit{any mis-aligned} signals. This issue can be further mitigated by \edit{building individual panels for either vertical or horizontal tuning only, and then interleaving those panels.}

\heading{Control circuitry.}
\edit{Each stepper motor is driven by a dedicated driver board~\cite{motor_drive_chip}, 
 which receives control signals from an ESP32 dev board~\cite{esp-32s}. }
%
\edit{We extend the ESP32 GPIOs with shift registers~\cite{shift-reg}, so that one ESP32 can simultaneously control 9 motors with 36 digital outputs total. The circuitry can be integrated into a PCB for robustness and ease future deployment. It is currently powered by 5V DC power supplies, but can use 
batteries instead for deployment flexibility. Given the precision of the stepper motors, a \name{} panel is able to control the rolling actuation at millimeter scale. We also perform a one-time calibration for motor control.}

\heading{Control infrastructure.} 
We control the panels with one controller (a Raspberry Pi 3B) and one edge server (a PC with an Intel I7-7700 processor) following \secref{sec:design}. 
\edit{The controller is set up as a network bridge, connecting to the server via Ethernet and to the four panels via 802.11n Wi-Fi. We implement the control algorithm in Matlab on the server, which can be easily moved to the controller. The control logic of every panel is implemented on the ESP32 dev board, which accepts commands for motor rotation.}
%
\edit{Our system architecture can also accommodate 
more than 4 panels and multiple controllers to improve coverage and performance.}


\heading{Hardware cost.}
Each panel comprises 9 sets of stepper motors and driver boards (costing around \$2.8 each), one ESP32 board (\$10), and 3 shift registers (\$0.28 each). The wooden structure for the panel costs around \$3. Therefore, each panel costs \$39. For our prototype with 4 panels, the total cost of \name{} is \$156. The hardware cost can be lowered further for large-scale production given economies of scale.



\heading{Hardware durability.}
The typical lifespan for stepper motors is around 5 years~\cite{motor-lifespan} and copper strips can be laminated with plastic sheets for durability. We can potentially fabricate the surfaces using a similar process to that for commercial flexible RFIDs. Note that our system has very light workload and only needs infrequent actuation, both of which improve the lifespan of the hardware. Further, as described in \secref{sec:design-control}, our algorithm makes few assumptions about the hardware. Even if the copper strips wear or deform slightly, our control algorithm can still explore and identify suitable surface configurations.

\section{Evaluation}
\label{sec:eval}

Our experiments revolve around assessing the unique feature of \name, i.e., frequency tunability and concurrent link support. We first validate the basic surface operations using single-link setups on different frequencies, in terms of RSSI and throughput gains. Then we highlight how \name\ effectively enhances concurrent links on multiple frequencies.

\subsection{Frequency Tunability Validation}
\label{sec:eval:freq-tunable}
\begin{figure}[t]
    \centering
    \includegraphics[width=\columnwidth]{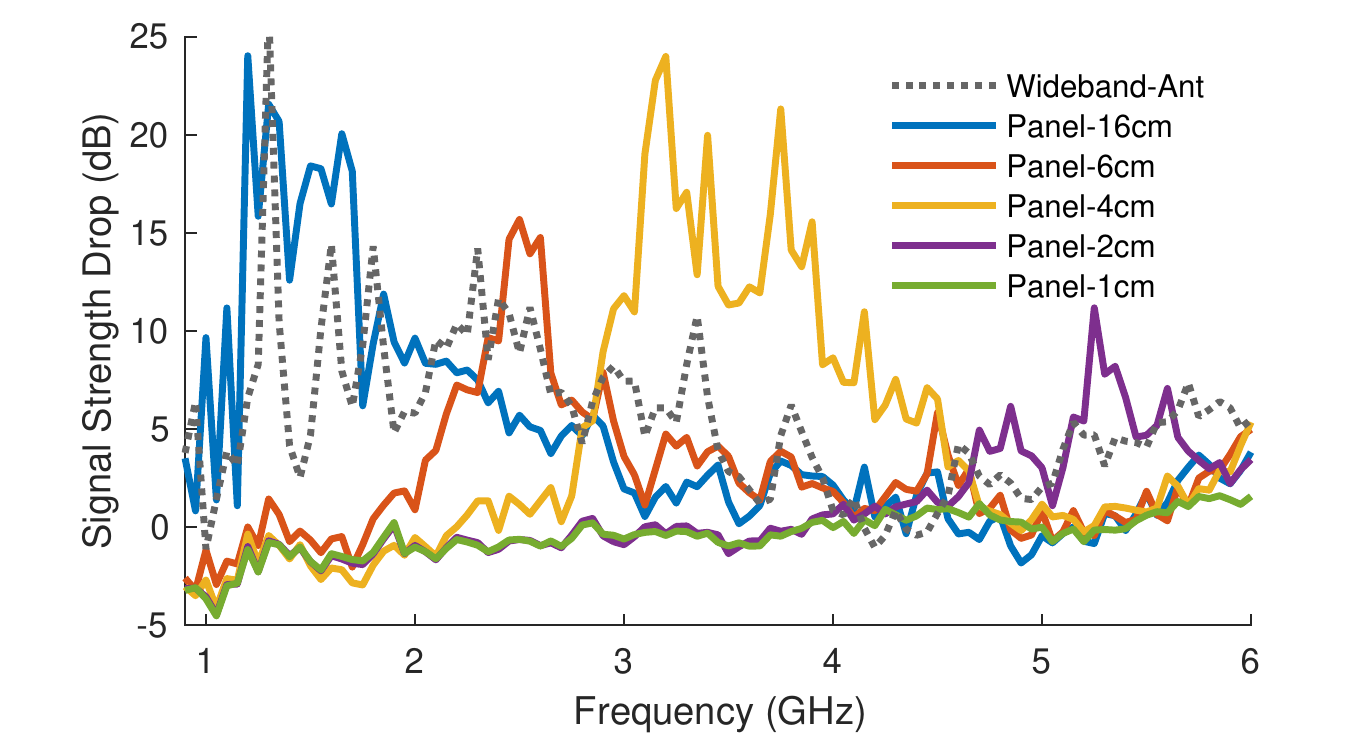}
    \caption{Signal strength changed by one \name\ panel, rolled to different lengths vs. a wideband antenna \edit{array}. }
    \label{fig:eval-freq-tunability}
 \end{figure}
The key feature of \name{} is to tune its operating frequency within sub-6~GHz and reflect the corresponding signals. To test the frequency tunability, we set up a 
short link with two USRP N210s, each connected to a directional Vivaldi antenna. The Vivaldi antennas are 1~meter apart and point towards each other. This distance roughly ensures we only measure the most direct, line-of-sight path between the antennas. We first measure the baseline received signal strength with nothing between the endpoints. Then we place one panel between the two USRPs and adjust the rolls to different lengths. The panel reflects signals and causes the received signal strength to drop on the resonant frequency, \myedit{i.e., peaks in \figref{fig:eval-freq-tunability}}. 
We see 
that the panel operates on the frequency ranges that matches its roll length. The maximum signal strength change is over 10~dB across all lengths, which means the panel reflects over 90\% of the incident signal power (only 10\% \myedit{transmitted through} for a 10~dB drop). When the rolls on the panel are 1~cm long (fully rolled), the received signal strength is very close to the \edit{no-surface} baseline. 
Thus, in the following sections, we use the measurements with 1~cm roll length as \textit{baselines}. 

To understand how our design compares with a wideband antenna array, we build a surface of the same size as the panel, affixed with commercial wideband antennas~\cite{flex600_antenna} as shown in \figref{fig:eval-wideband-surf}. The wideband antennas cover 
cellular and ISM frequencies from 0.6--6~GHz. Due to the large antenna size, we cannot know the best spacing for the wideband antennas or choose the same 3~cm spacing as for \name. 
Therefore, we arrange the antennas with a relatively dense 6~cm spacing. \figref{fig:eval-freq-tunability} shows the signal strength change caused by the wideband antennas is non-tunable in the frequency domain and much smaller than what \name\ can achieve. 

\begin{figure}[t]
  \begin{subfigure}{0.95\columnwidth}
    \centering
    \includegraphics[width=0.95\columnwidth]{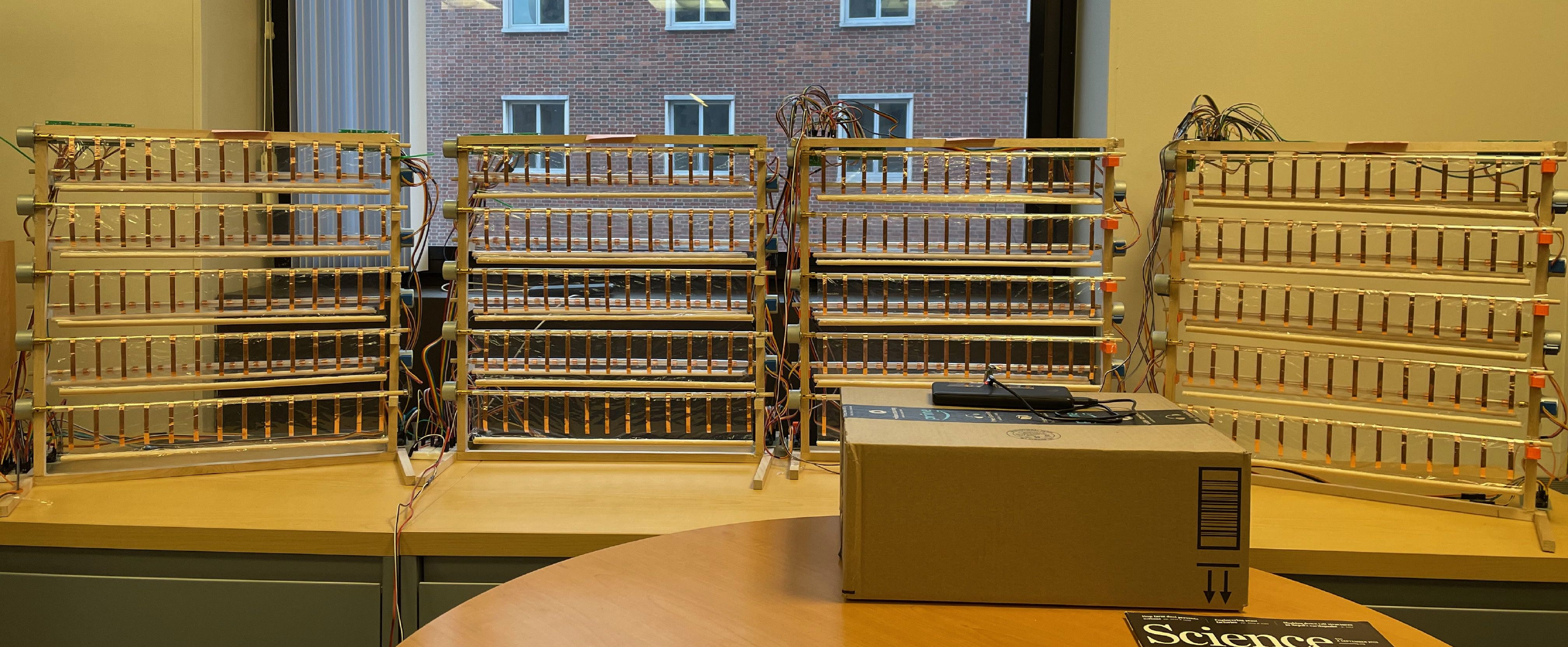}
    \caption{Setup\#1: Multiple panels in a row.}
    \label{fig:eval-panels-setup1}
  \end{subfigure}
  \bigskip\\
  \begin{subfigure}{0.47\columnwidth}
    \centering
    \includegraphics[width=0.95\columnwidth]{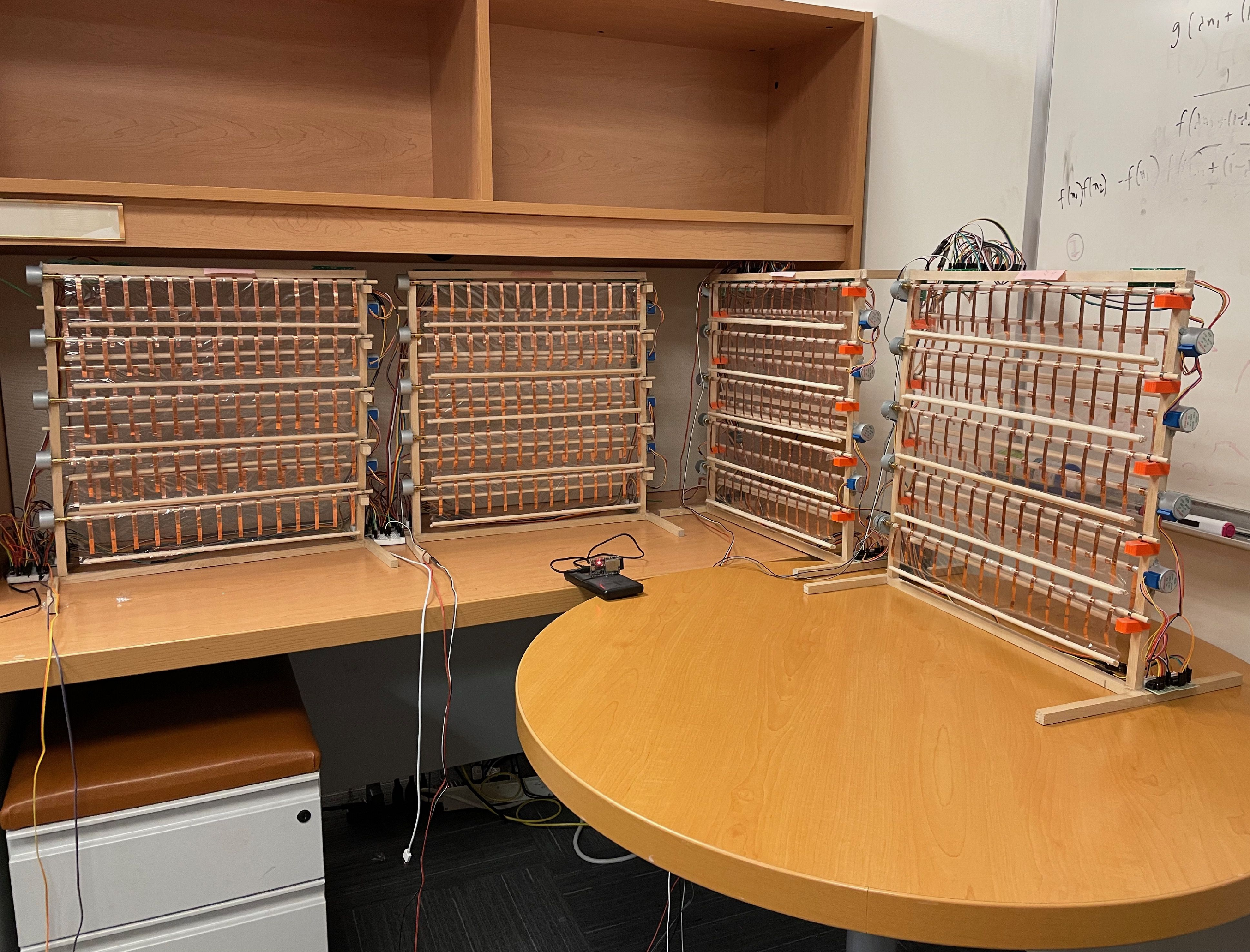}
    \caption{Setup\#2: Perpendicular.}
    \label{fig:eval-panels-setup2}
  \end{subfigure}%
  \begin{subfigure}{0.47\columnwidth}
    \centering
    \includegraphics[width=0.95\columnwidth]{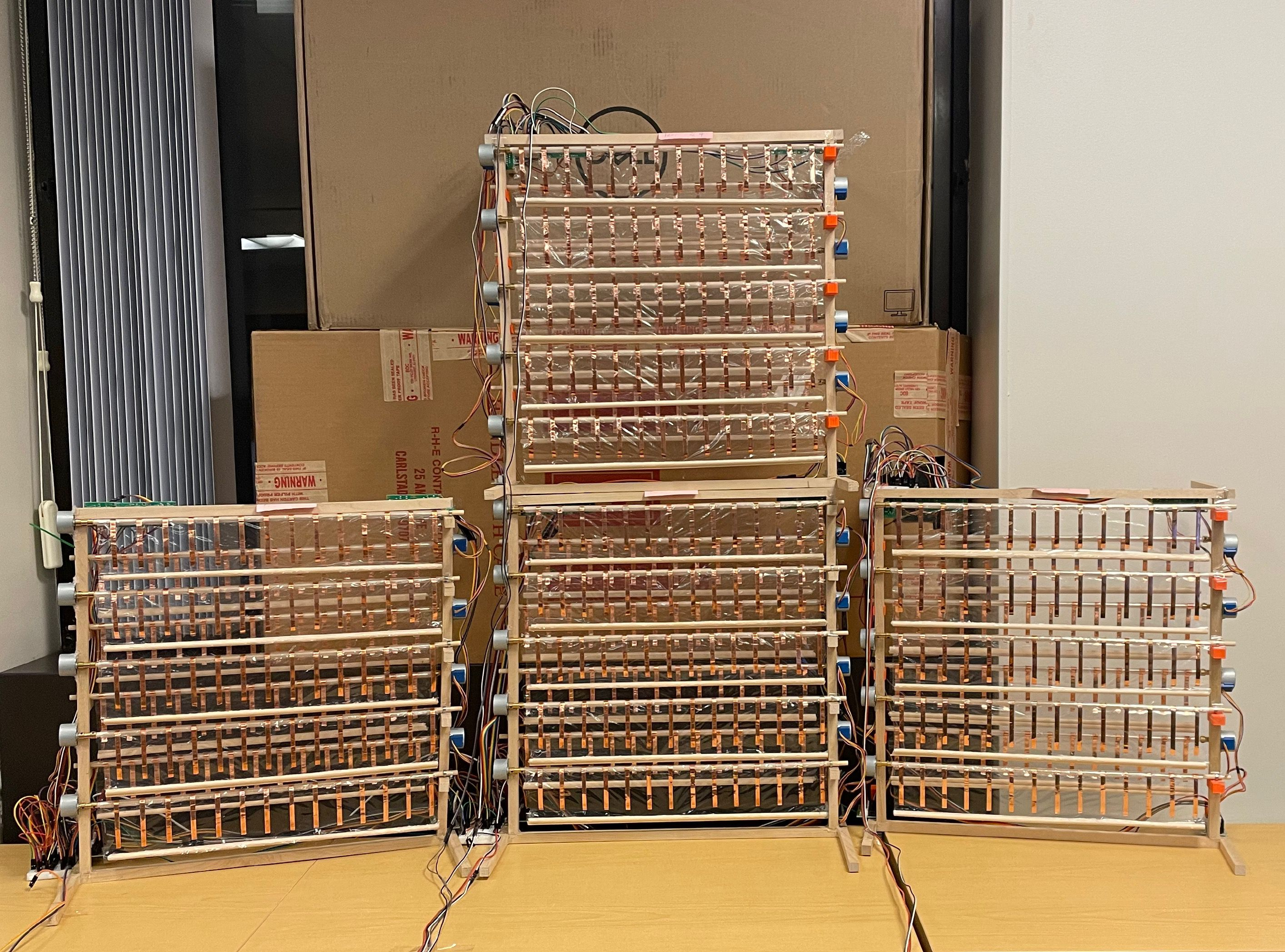}
    \caption{Setup\#3: Stacked up.}
    \label{fig:eval-panels-setup3}
  \end{subfigure}%
  \caption{Panel Setups.}
  \label{fig:eval-panels-setup}
\end{figure}

\begin{figure}[t]
  \centering
  \includegraphics[width=0.9\columnwidth]{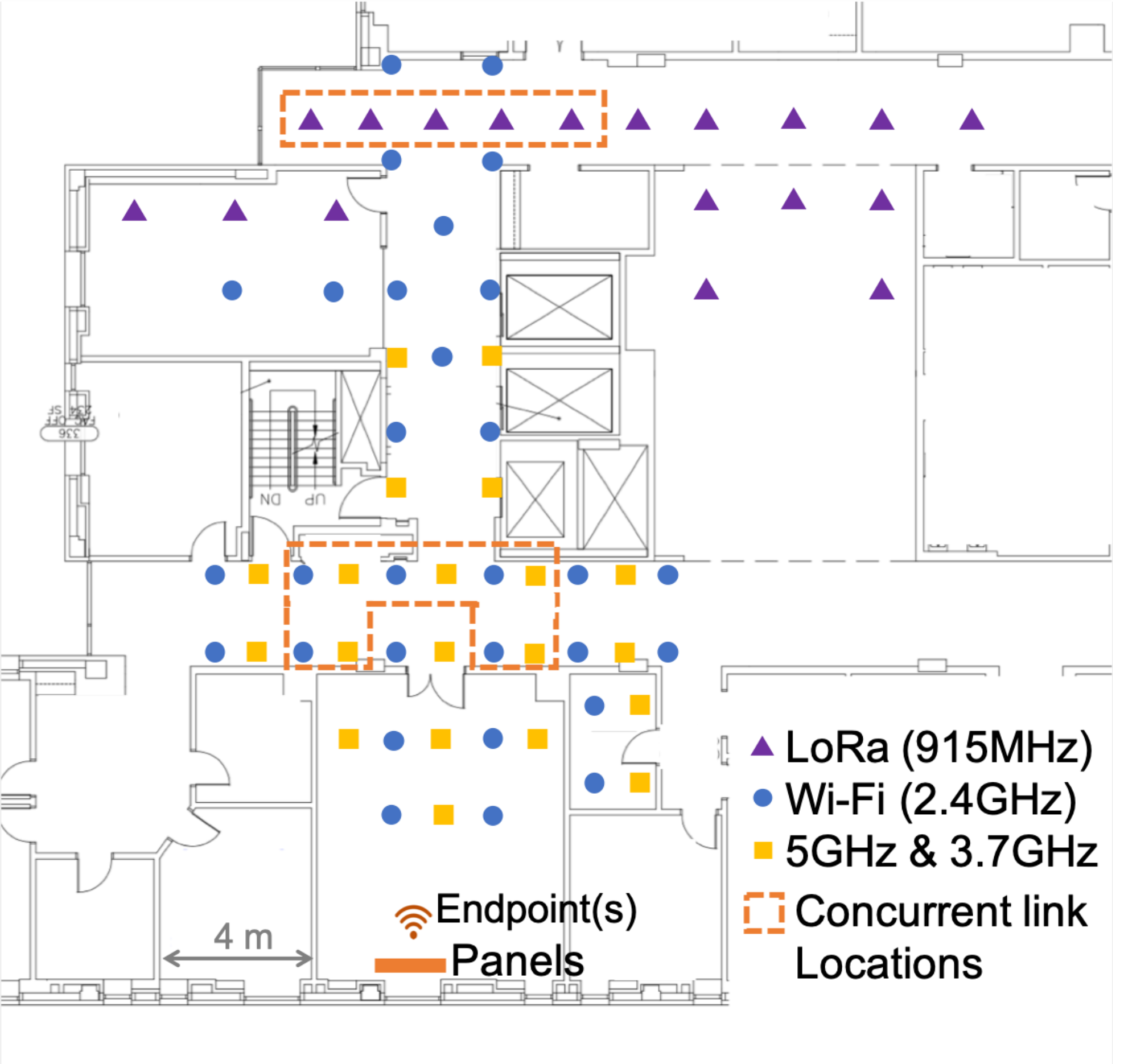}
  \caption{Link Locations.}
  \label{fig:eval-links-map}
\end{figure}


\subsection{Large-Scale Experiment Setup}
For large-scale experiments, we deploy \name{} prototypes on the desk in an office (\figref{fig:eval-panels-setup}). There are 14 other desks in this office, with tall bookshelves and objects on the desks. We use setup\#1(\figref{fig:eval-panels-setup1}) and place the wireless endpoint(s) 35~cm in front of the panels 
unless otherwise stated. The other endpoint is moved around the floor to sample different channel conditions. The link locations for the following experiments are marked in \figref{fig:eval-links-map}. This mimics a typical application scenario where a user has multiple wireless devices on the desk, for example, a laptop connected to Wi-Fi, a phone connected to the nearest cellular base station, and IoT devices connected to Wi-Fi or LoRa. These devices may operate on the same frequency band, or, more likely, on different frequencies. \name{} is not limited to the desk placement, nor to a specific panel setup. We envision that \name{} panels can be placed anywhere close to the wireless endpoints or be integrated with building structures or furniture. 

We deploy diverse wireless links for evaluation. Specifically, we deploy 915~MHz LoRa links using two LoRa boards\cite{esp32-lora} and 2.4~GHz 802.11n Wi-Fi links using ESP32 devkit-c boards~\cite{esp32-devkitc}, on channel 1 with 40~MHz bandwidth. The receivers send the RSSI feedback values to the sub-controller using Wi-Fi. We also set up links on the 5.21~GHz Wi-Fi band and the 3.7~GHz 5G cellular C band with WARPv3 devices or USRP N210s~\cite{usrp-n210}, using VERT2450 antennas~\cite{vert2450} and wideband cellular antennas~\cite{flex600_antenna} for the respective bands. The transmitter sends a 1~MHz bandwidth square wave on the test frequency; the receiver reports received signal strength to the controller over Ethernet. We place the receiver in front of panels unless otherwise stated.
\edit{These frequencies are selected due to device availability, and we expect to see similar enhancement from \name\ on other sub-6~GHz bands. }

\begin{figure}[t]
  \centering
  \includegraphics[width=0.95\linewidth]{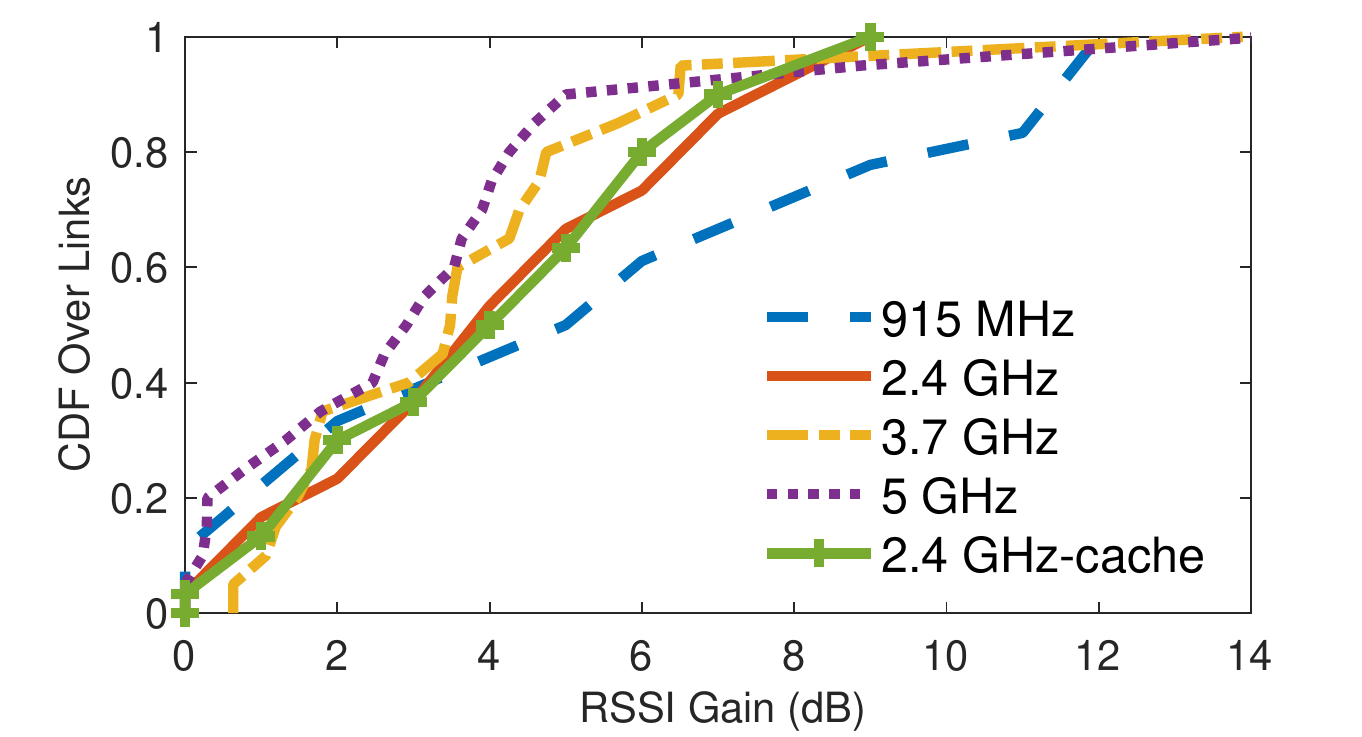}
  \caption{RSSI gain for individual links on different frequencies and with cached \name\ configurations.}
  \label{fig:eval-single-link-gains}
\end{figure}

\begin{figure*}[t]
  \begin{minipage}{.32\textwidth}
    \centering
    \includegraphics[width=\linewidth]{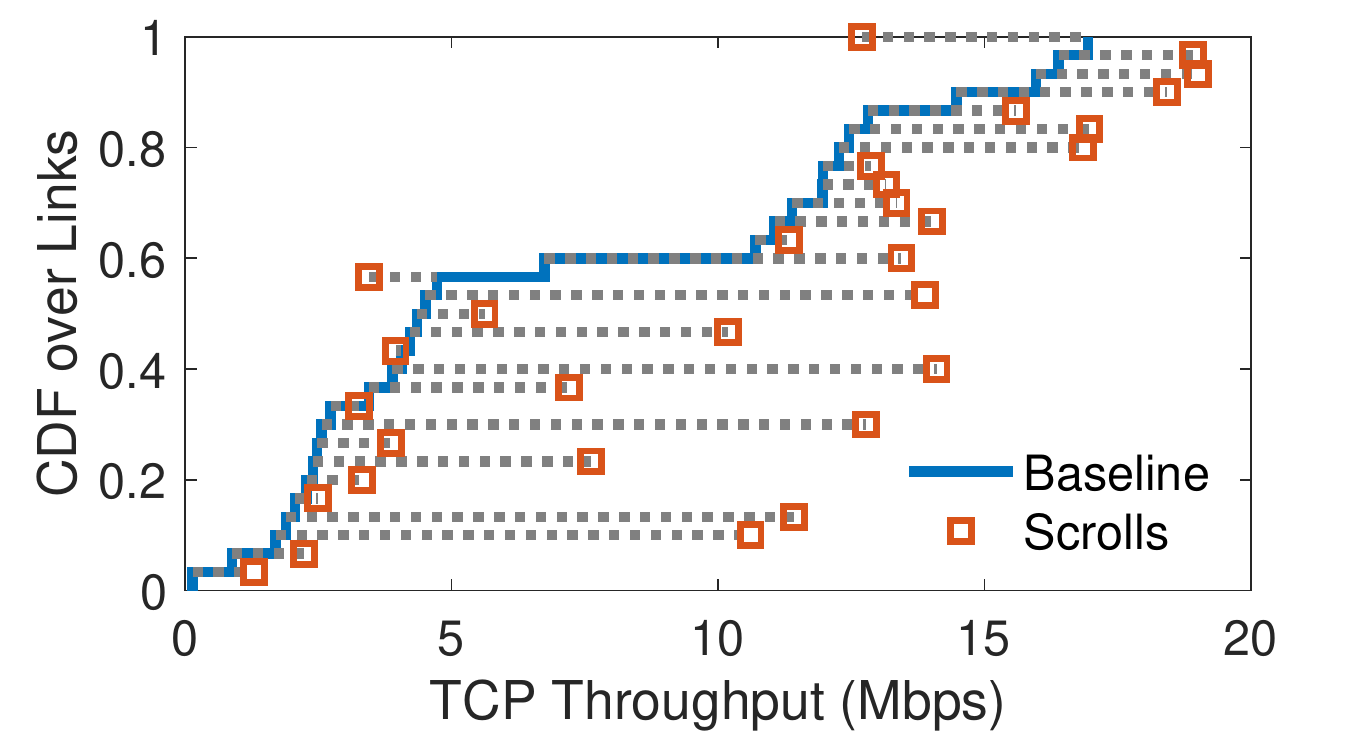}
    \caption{TCP throughput improvement for 2.4~GHz Wi-Fi links.}
    \label{fig:eval-esp-wifi-tcp-cdf}
  \end{minipage}\hfill
  \begin{minipage}{.32\textwidth}
    \centering
    \includegraphics[width=\columnwidth]{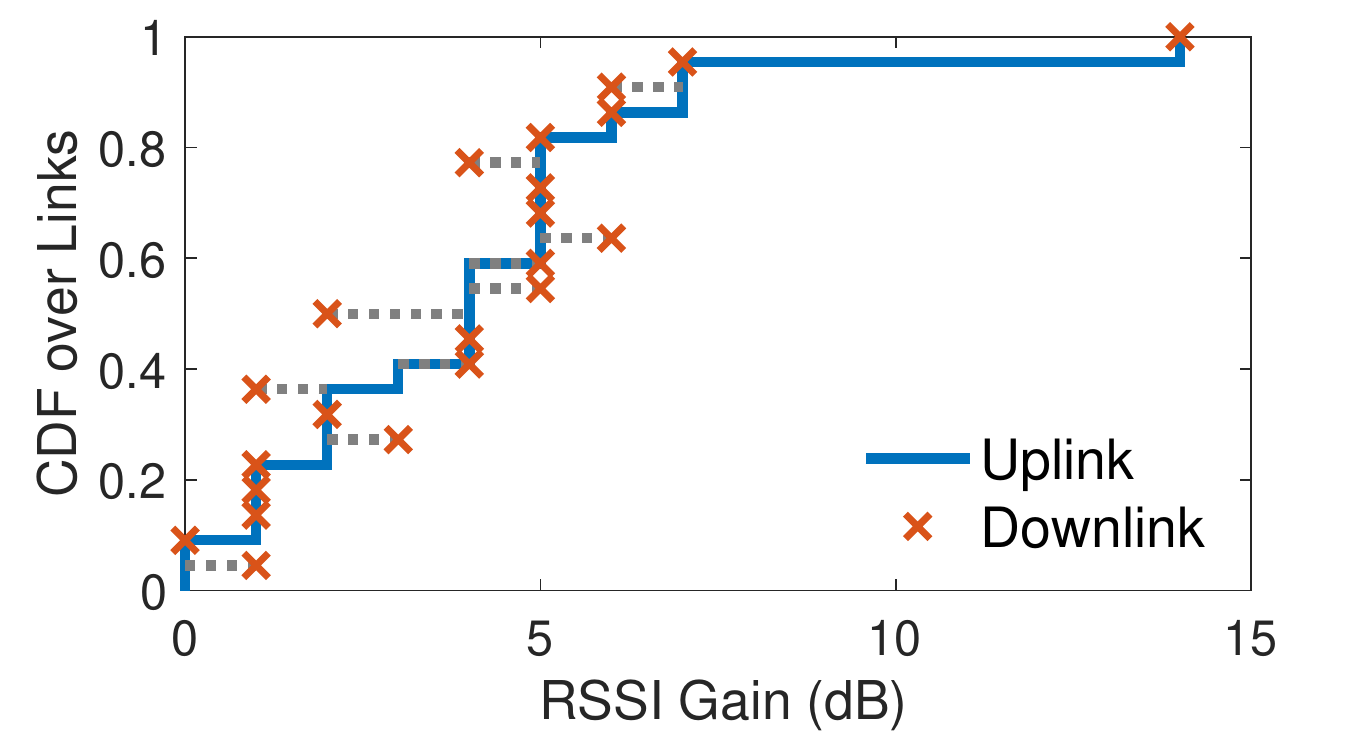}
    \caption{RSSI gains for Wi-Fi \\ uplink/downlink pairs on 2.4~GHz. }
    \label{fig:eval-esp-wifi-reciprocity}
  \end{minipage}\hfill
  \begin{minipage}{.32\textwidth}
    \centering
    \includegraphics[width=\columnwidth]{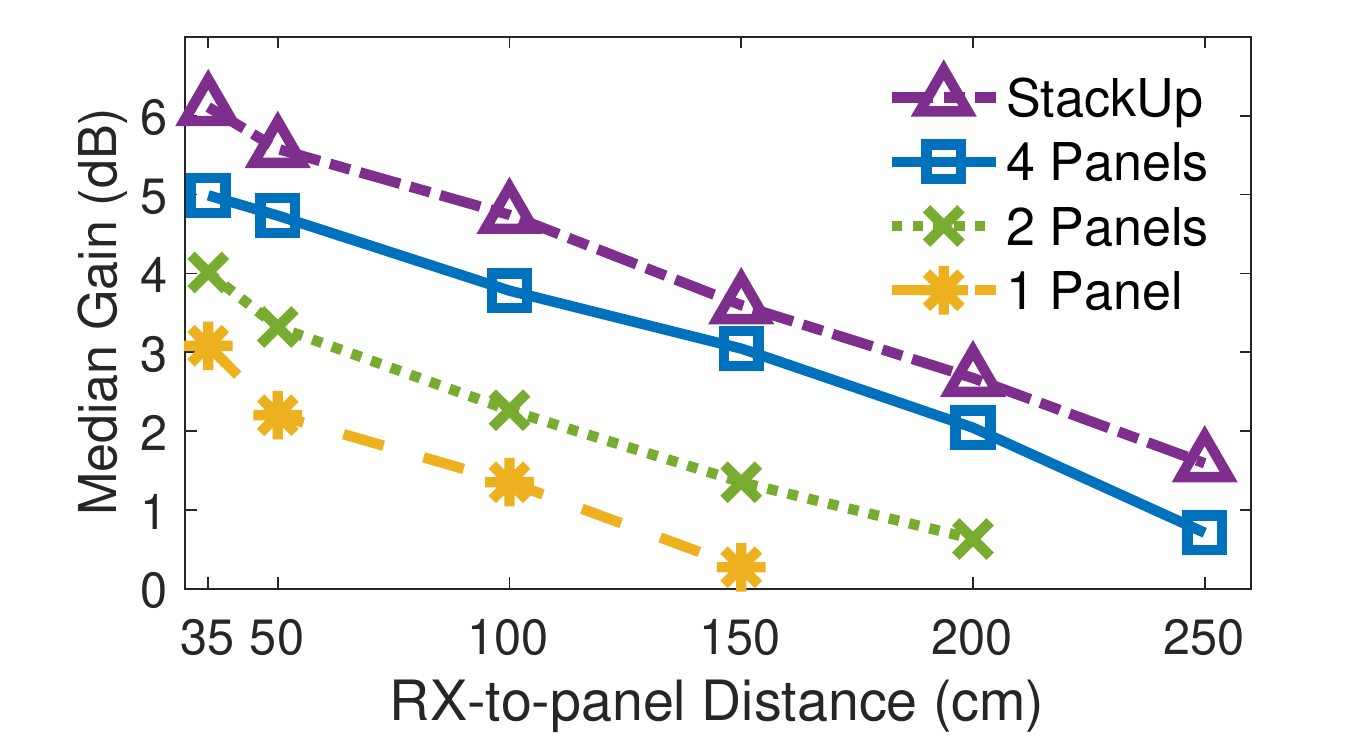}
    \caption{Median signal strength gain versus RX-to-panel distance.}
    \label{fig:eval-gain-over-distance}
  \end{minipage}
\end{figure*}


\subsection{Enhancing Single Links}
%
\figref{fig:eval-single-link-gains} presents the received signal strength gains from \name{} for different frequencies. 
For 900~MHz, we set up 18 LoRa links, placing the transmitter relatively far from the receiver and the panels, 
given LoRa devices are primarily used for long-range communication. 
%
\name{} can provide a median RSSI gain of 5.5~dB and up to 12~dB. This potentially enables LoRa devices to use a lower spreading factor for high bitrate or have a larger communication range. Although LoRa devices are typically used for outdoor applications, our indoor experiments demonstrate the benefit of \name{} on 900~MHz.
For 2.4~GHz, we setup 30 Wi-Fi 802.11n links, uniformly sampling the communication range of ESP32 devices. The results show a median gain of 4~dB and a maximal gain of 9~dB. As described in \secref{sec:cache-scheme}, \name{} can cache the surface configuration for a link. 
\figref{fig:eval-single-link-gains} further shows the cached surface configurations provide a similar enhancement. 
For 3.7~GHz and 5.21~GHz, with 20 links each, \name\ provides a median gain of 3.5~dB for 3.7~GHz and 3~dB for 5.21~GHz; the maximal gain is around 14~dB for both frequencies. 

To summarize, \name{} can improve the received signal strength for links on the 915~MHz, 2.4~GHz, 3.7~GHz and 5~GHz bands, and we expect this to apply to other sub-6~GHz frequencies as well. 
The median RSSI \edit{improvement} decreases as the frequency gets higher. This happens for two reasons. First, according to Friis' equation, the ratio between the power of the direct path and the reflected path is frequency-dependent; It decreases with increasing frequency, as calculated in ScatterMIMO~\cite{scattermimo}. Second, \name{} currently fixes the spacing between \edit{vertically fitted rows; the exposed row length decreases as the operating frequency increases, and the overall reflective antenna array becomes sparser.} 


\heading{Throughput increase.}
To show the throughput increase brought by the RSSI gain, we run \texttt{iperf} measurements of TCP and UDP for 30~sec on 2.4~GHz Wi-Fi links, with and without \name. 
\figref{fig:eval-esp-wifi-tcp-cdf} shows that the mean throughput increase is 47\% for TCP and 48.4\% for UDP, while the maximal throughput increase is around 8$\times$ for TCP and 7$\times$ for UDP. Two links incur lower throughput after activating \name{}. Aside from measurement noise, this is possibly due to several OFDM subcarriers undergoing substantial SNR drop, which leads to a lower throughput even though the aggregate RSSI increases slightly. 
We can address this issue by leveraging subcarrier-level channel state information in the control algorithm.

\heading{Other link and panel setups.} 
Although we place the receiver near the panels and use panel setup\#1 (\figref{fig:eval-panels-setup1}) by default, \name{} works for other link and panel setups as well. We set up 20 Wi-Fi links with setup\#2 (\figref{fig:eval-panels-setup2}) and achieve a median gain of 4~dB in both link directions (\figref{fig:eval-esp-wifi-reciprocity}). The endpoint near the panels acts as the receiver (RX) for uplink or as the transmitter (TX) for downlink. \name{} provides similar gains to both directions using only uplink feedback, i.e., the enhancement applies to both directions alike, due to the reciprocity of wave propagation. Therefore, \name{} can be placed near either the receiver or the transmitter.

\heading{Panel coverage.} 
Similar to existing work on (passive) smart surfaces, \name{} should be close to at least one endpoint of the link. To show the panel coverage, we vary the distance between RX to the nearest panel, from 35~cm to 2.5~m, while placing TX around 10~m away from the panels. For each distance, we take 15 measurements on 2.4~GHz and show the median RSSI gain to avoid variations caused by multipath fading. We also compare the performance for different panel layouts 
(\figref{fig:eval-panels-setup3}). As shown in \figref{fig:eval-gain-over-distance}, \name{} (4 panels) can provide a median gain of over 3~dB, with the RX 150~cm from the panels.
At any given distance, more panels improves the median gain; for a given performance target, more panels increases the coverage area. 
A denser panel layout (e.g., stacked up) further improves the performance.

\heading{Comparison with other smart surfaces.}
The frequency-tunability goal of \name\ sets it apart from existing, single-band surfaces, which generally focus on leveraging different signal manipulation mechanisms. Further, the \name\ prototype follows a different deployment strategy, and hence control mechanism, than existing prototypes.  
Therefore, we qualitatively compare \name\ to others in terms of single-band performance.
RFocus~\cite{rfocus} is the most similar to \name\ in terms of signal manipulation and provides a median gain of 9.5~dB with a surface size of 6~$m^2$ on 2.4~GHz, while \name{} provides a median gain of 4~dB with 4 panels, only 0.2~$m^2$ each. If the gain scales based on $10*log(N)$~\cite{beyond-massive-mimo} (where $N$ is the element count), \name{} could provide 12.7~dB median gain with a 6~$m^2$ surface area; assuming $10*log(N^2)$ scaling~\cite{rfocus}, \name{} could provide 21.5~dB gain with 6~$m^2$.
LAVA~\cite{lava-sigcomm21} shows higher gains, \edit{but operates as a multi-hop amplify-and-forward mesh, using active power amplifiers, whereas \name\ employs single-hop (passive) signal reflection.}
ScatterMIMO~\cite{scattermimo} is specific to MIMO APs, while \name{} currently targets single-antenna devices.
RFlens~\cite{RFlens} assumes a short (3~m) link with only a \textit{transmissive} surface in between \edit{endpoints} to achieve the best performance.
Overall, \name{} achieves comparable gains using much simpler, low-cost hardware, and a relatively small surface area.


\begin{figure}[t]
  \centering
\begin{subfigure}{0.49\columnwidth}
  \centering
  \includegraphics[scale=0.3]{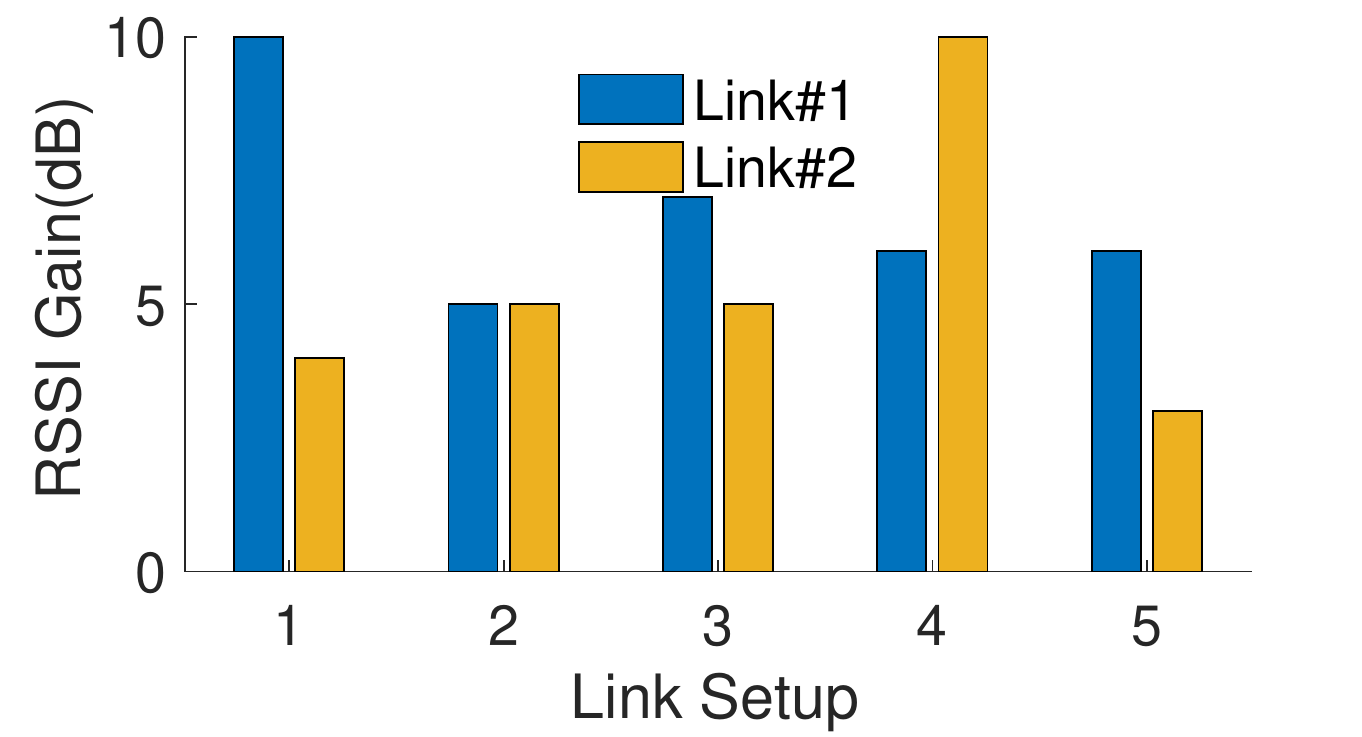}
  \caption{RSSI Gain.}
  \label{fig:eval-esp-wifi-2links-rssi-gain}
\end{subfigure}%
\begin{subfigure}{0.49\columnwidth}
  \centering
  \includegraphics[scale=0.3]{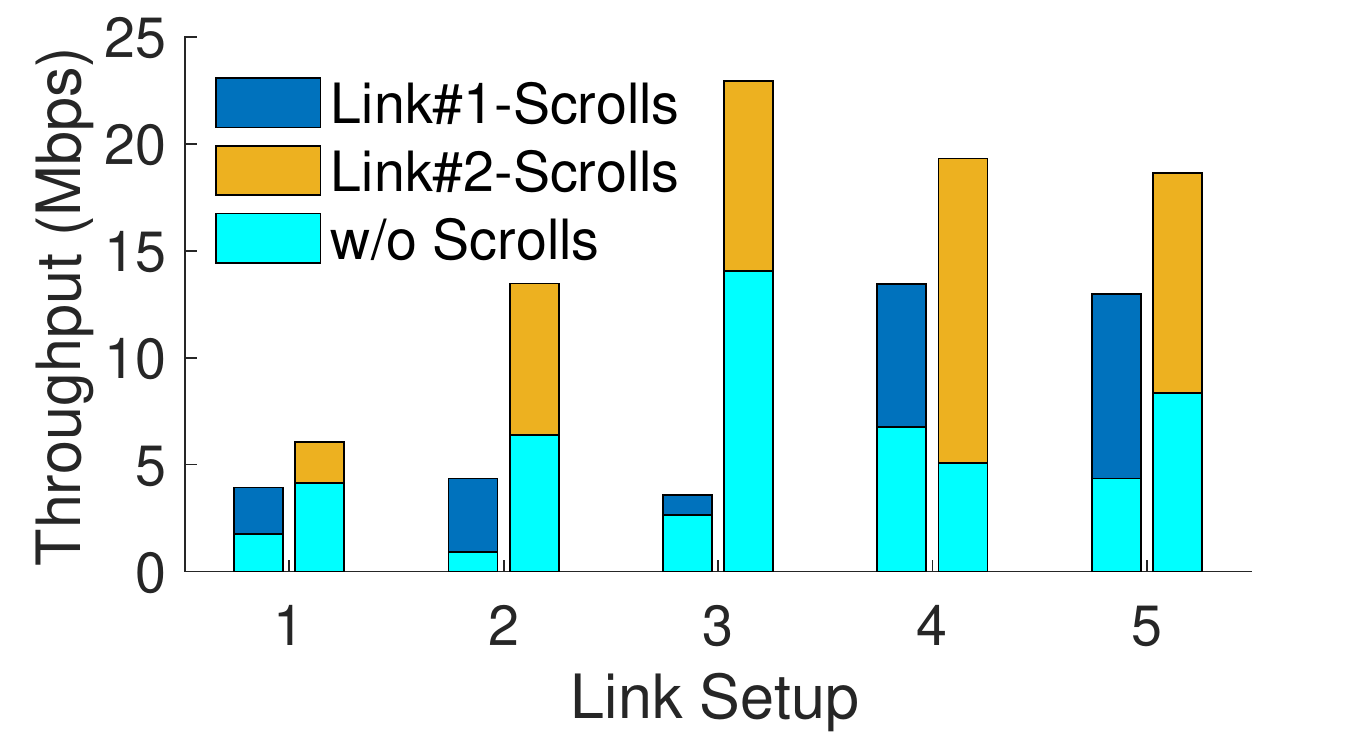}
  \caption{TCP Throughput.}
  \label{fig:eval-esp-wifi-2links-tcp}
\end{subfigure}
\caption{Enhancing two 2.4~GHz Wi-Fi links.} 
\label{fig:eval-esp-wifi-2links}
\end{figure}


\subsection{Enhancing Multiple Links}
We next evaluate \name{}' performance when there are multiple links on the same or different frequency bands.

\heading{Two Wi-Fi links.}
We test 
\edit{pairs} of ESP32 Wi-Fi AP-station links on the same frequency with 5 different link setups. We place the receivers in front of the panels, and move the transmitters to different locations. Only one link can actively transmit at any given time since they share a channel. \name{} does not know which link is active, so it configures the panels to provide enhancement to both links. As shown in \figref{fig:eval-esp-wifi-2links-rssi-gain}, \name{} provides a median gain of 5.5~dB among all links and a maximum of 10~dB. \figref{fig:eval-esp-wifi-2links-tcp} shows the corresponding TCP throughput for each link with and without \name{}. The median TCP throughput increase is 110\% and the maximum increase is 378\%. Thus, \name{} is able to support multiple users on the same frequency in close proximity.

\begin{figure*}[t]
  \centering
  \begin{subfigure}{0.33\linewidth}
    \centering
    \includegraphics[width=\columnwidth]{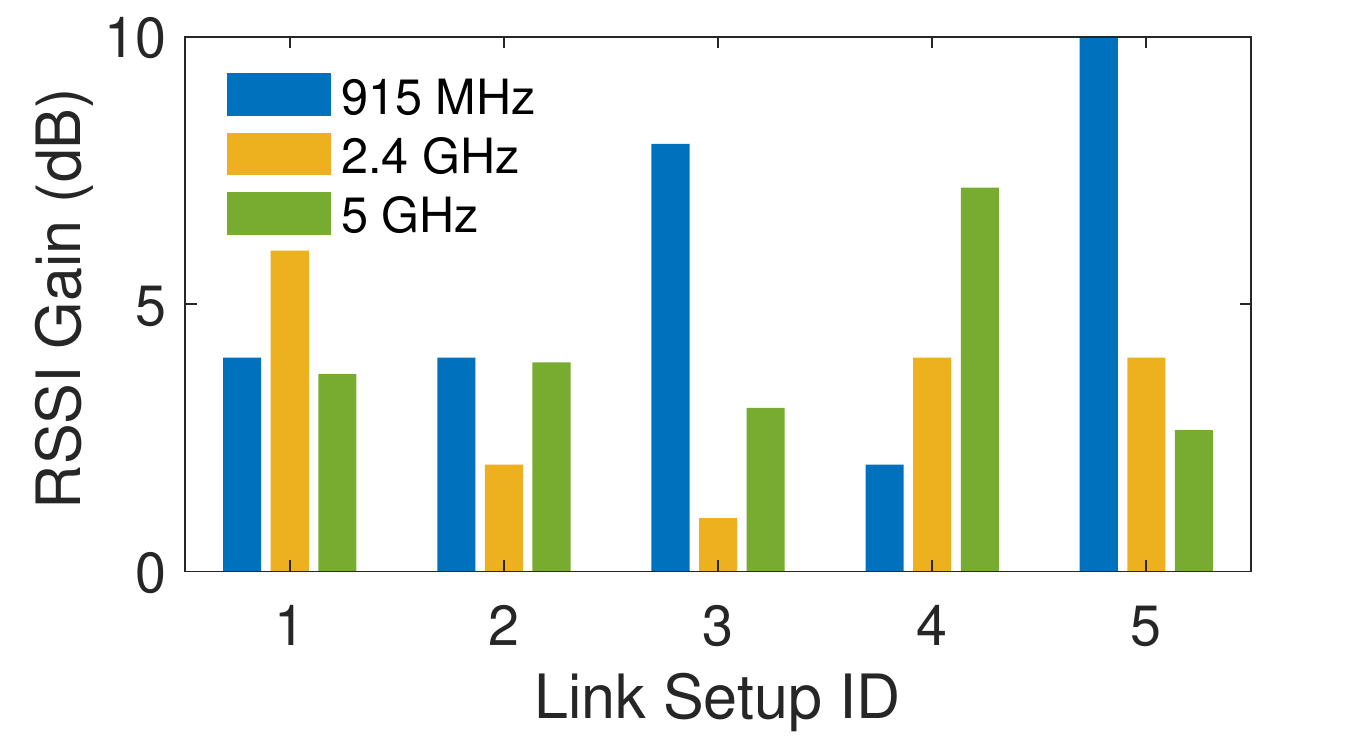}
    \caption{\textmd{3 concurrent links.}}
    \label{fig:eval-conc-3links-rssi-gain}
  \end{subfigure}
  \begin{subfigure}{0.33\linewidth}
    \includegraphics[width=\columnwidth]{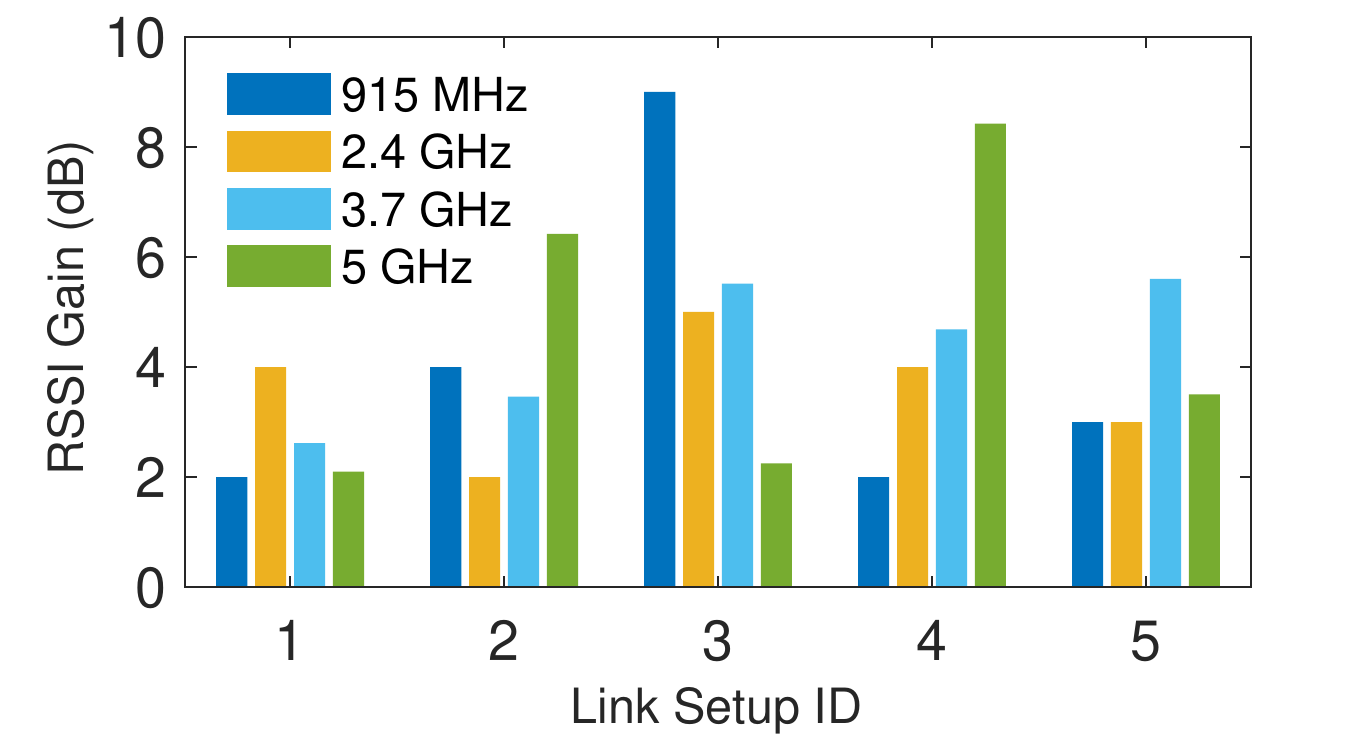}
    \caption{\textmd{4 concurrent links.}}
    \label{fig:eval-conc-4links-gain-1}
  \end{subfigure}
  \begin{subfigure}{0.33\linewidth}
    \includegraphics[width=\linewidth]{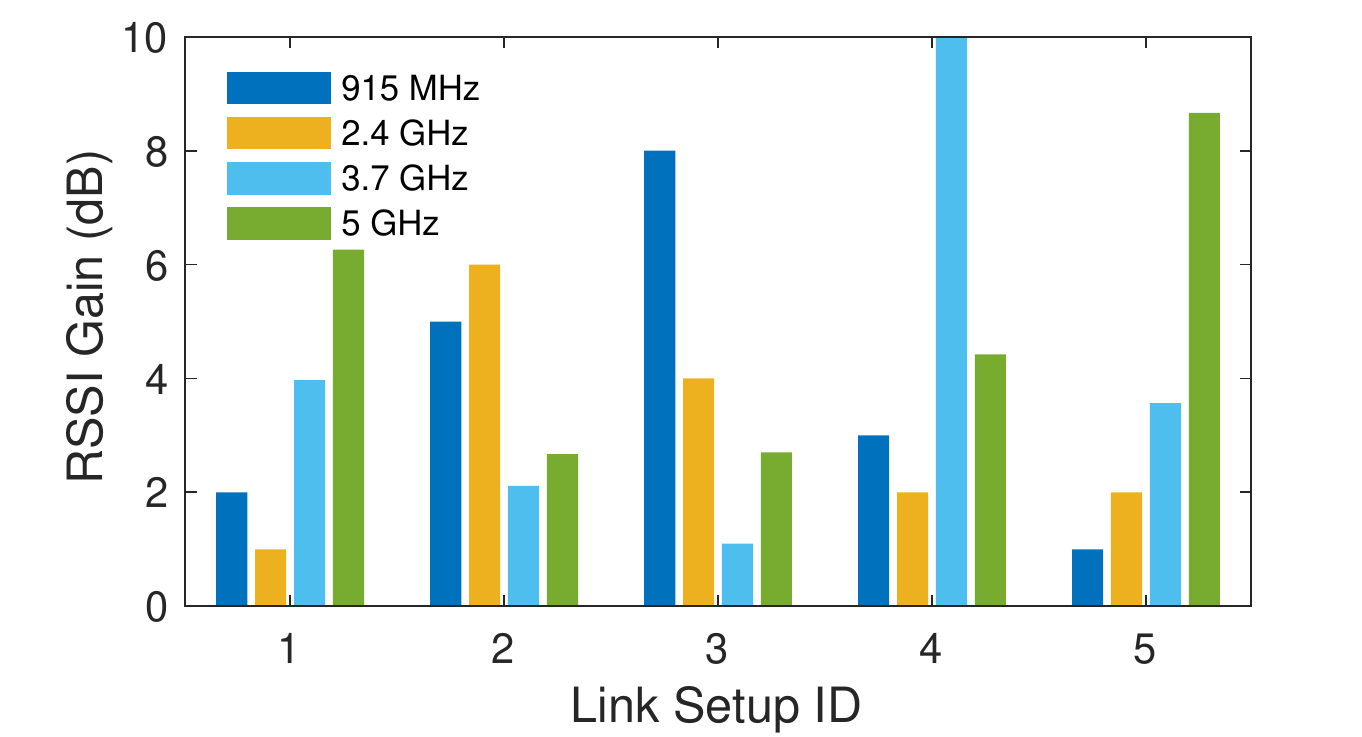}
    \caption{\textmd{4 links with RX endpoints stacked closely.}}
    \label{fig:eval-conc-4links-gain-2}
  \end{subfigure}
  \caption{\name{} enhances concurrent links on different frequencies, with a median gain of 4~dB and up to 10~dB.}
  \label{fig:eval-conc-links}
\end{figure*}

\heading{Multiple concurrent links.}
We next \myedit{group multiple concurrent links on different frequencies, specifically on 915~MHz (LoRa), 2.4~GHz (Wi-Fi), 3.7~GHz, and 5~GHz.}
We place receivers \edit{one each} in front of each panel and move the transmitters around for 5 link \edit{group} setups (\figref{fig:eval-links-map}). \myedit{We measure RSSI gain for group setups containing three links (\figref{fig:eval-conc-3links-rssi-gain}) and four links (\figref{fig:eval-conc-4links-gain-1}). \name{} provides a median RSSI gain of 4~dB and up to 10~dB across links of all group setups.} 
We then stack the 4 receivers close together in a $2\times2$ grid, with a 20~cm vertical separation and a 30~cm horizontal separation. It is more challenging to provide gains to all endpoints in this case (\figref{fig:eval-conc-4links-gain-2}), since the endpoints share the same set of nearest surface elements. \name{} still provides a median gain of over 3~dB and up to 10~dB.

\heading{Comparison with other smart surfaces.}
\edit{LAIA~\cite{laia-nsdi19} can enhance two concurrent links on \textit{different} channels on 2.4~GHz. LAVA~\cite{lava-sigcomm21} further supports two concurrent links on the \textit{same} channels. However, both were designed for \textit{a single frequency band}. } 
In contrast, \name{} provides comparable received signal strength and throughput gains for multiple links on \textit{either the same or different sub-6~GHz bands} simultaneously.


\subsection{Dynamics of Multiple Panels} 
\label{sec:eval:multi-panel}
\begin{figure}[t]
  \begin{minipage}{.45\columnwidth}
    \centering
    \includegraphics[width=\linewidth]{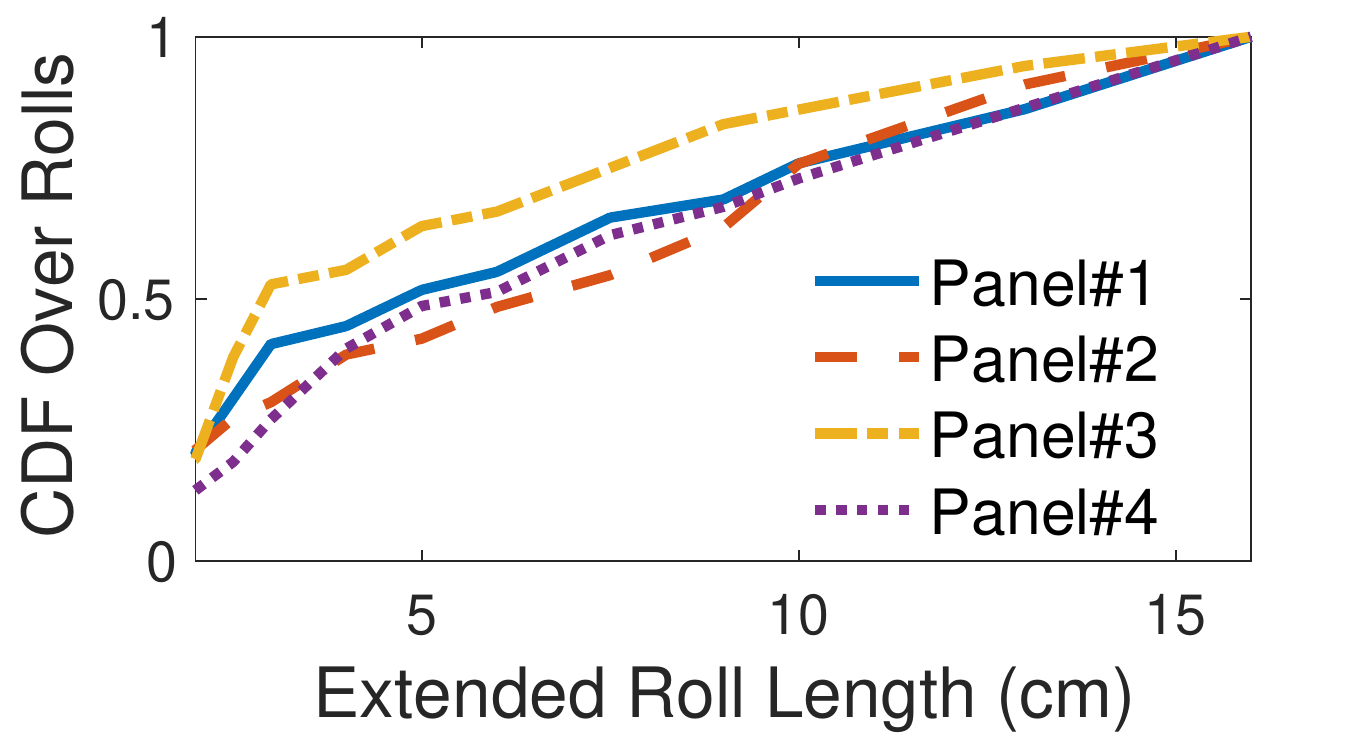}
    \caption{Extended roll length distributions.}
    \label{fig:eval-rlens-cdf}
  \end{minipage}\hfill
  \begin{minipage}{.48\columnwidth}
    \centering
    \includegraphics[width=\linewidth]{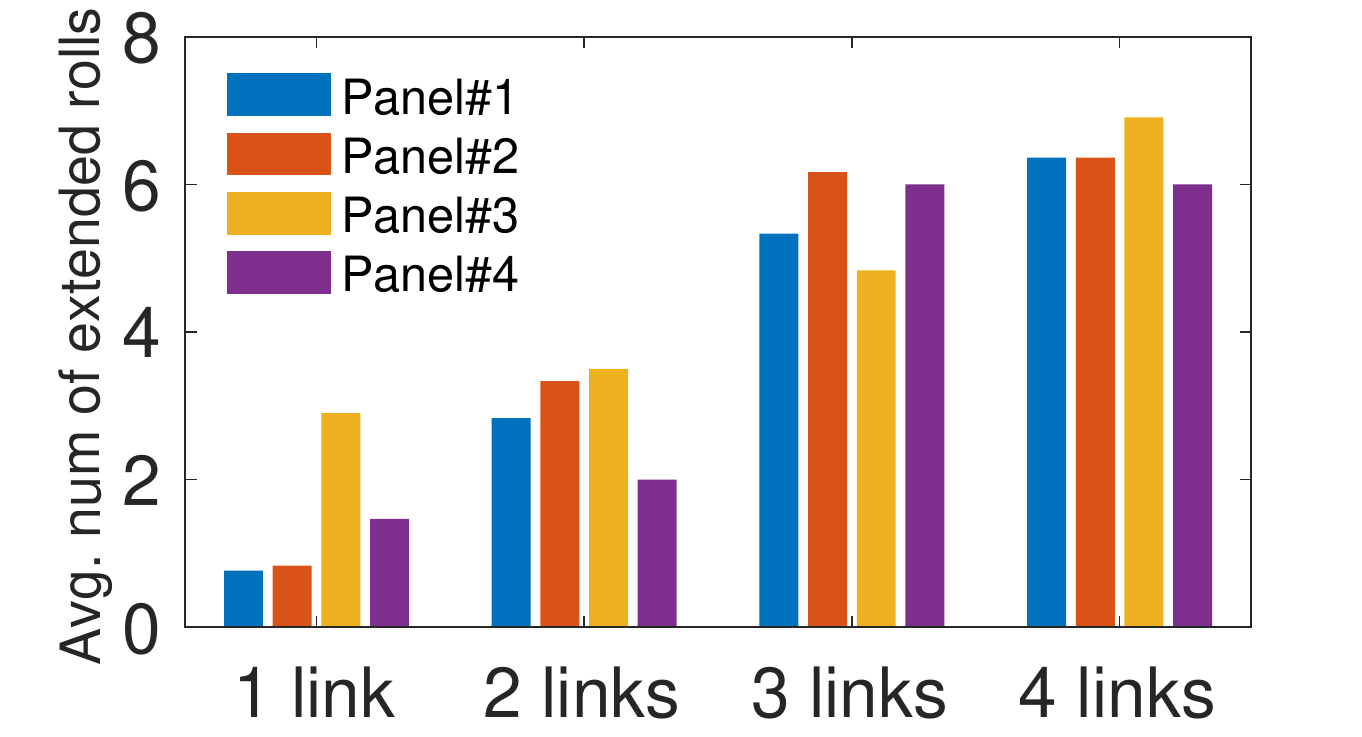}
    \caption{Dynamics of multiple panels.}
    \label{fig:eval-on-ele-per-panel}
  \end{minipage}
\end{figure}

We next analyze the dynamics of multiple composable \name\ panels.
We compare the distribution of roll lengths on each panel, using results from \figref{fig:eval-conc-3links-rssi-gain} and \figref{fig:eval-conc-4links-gain-1}. The distributions for each of the 4 panels (\figref{fig:eval-rlens-cdf}) are very similar \edit{regardless of the frequency of the receiver closest to them}. 
This highlights the necessity and advantages of \edit{\textit{intra-surface} (roll-wise currently)} frequency tunability. Different rolls can be dynamically assigned to distinct frequencies, \edit{individualized for links}. 

Using the optimized roll states of links in previous experiments, we also plot the average numbers of extended rolls per panel (\figref{fig:eval-on-ele-per-panel}). 
For one \edit{endpoint} 
placed in front of panel\#3, panel\#3 shows a much higher number of extended rolls compared to the other three panels, \edit{delivering most of the gain}. 
However, the other panels are still necessary to collectively increase the coverage of the system and the performance. For two concurrent links, two panels show over three extended rolls on average, while the other two panels also unfurled significantly. For three and four concurrent links, all panels have a sufficient number of rolls extended. In this case, multiple panels need to work together to provide sufficient gains to all links. It is very likely that in this setup, the gain provided by our system is upper bounded by the limited number of panels. 

\begin{figure*}[t]
  \begin{minipage}{.33\linewidth}
    \centering
    \includegraphics[width=\linewidth]{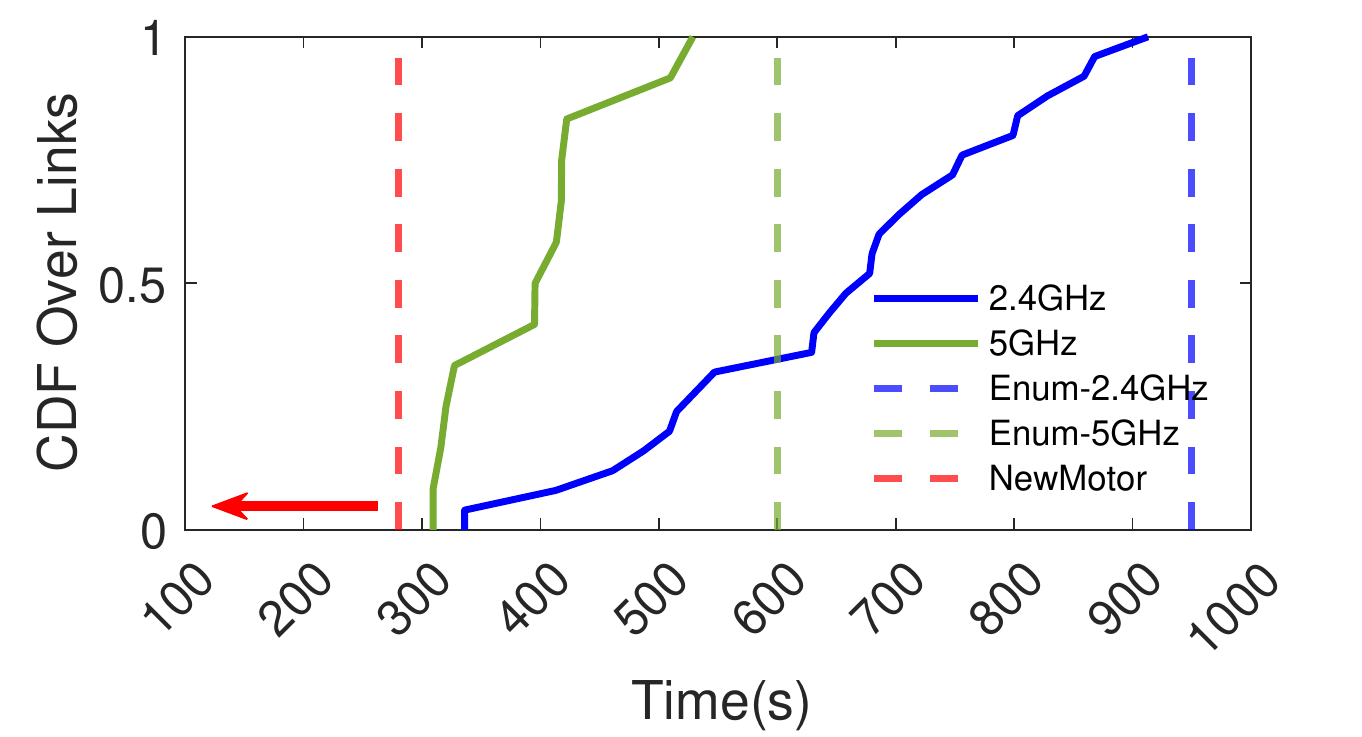}
    \caption{Algo convergence time.}
    \label{fig:eval-scrolls-opt-time}
  \end{minipage}\hfill
  \begin{minipage}{.64\linewidth}
    \begin{subfigure}{0.48\columnwidth}
      \centering
      \includegraphics[width=\linewidth]{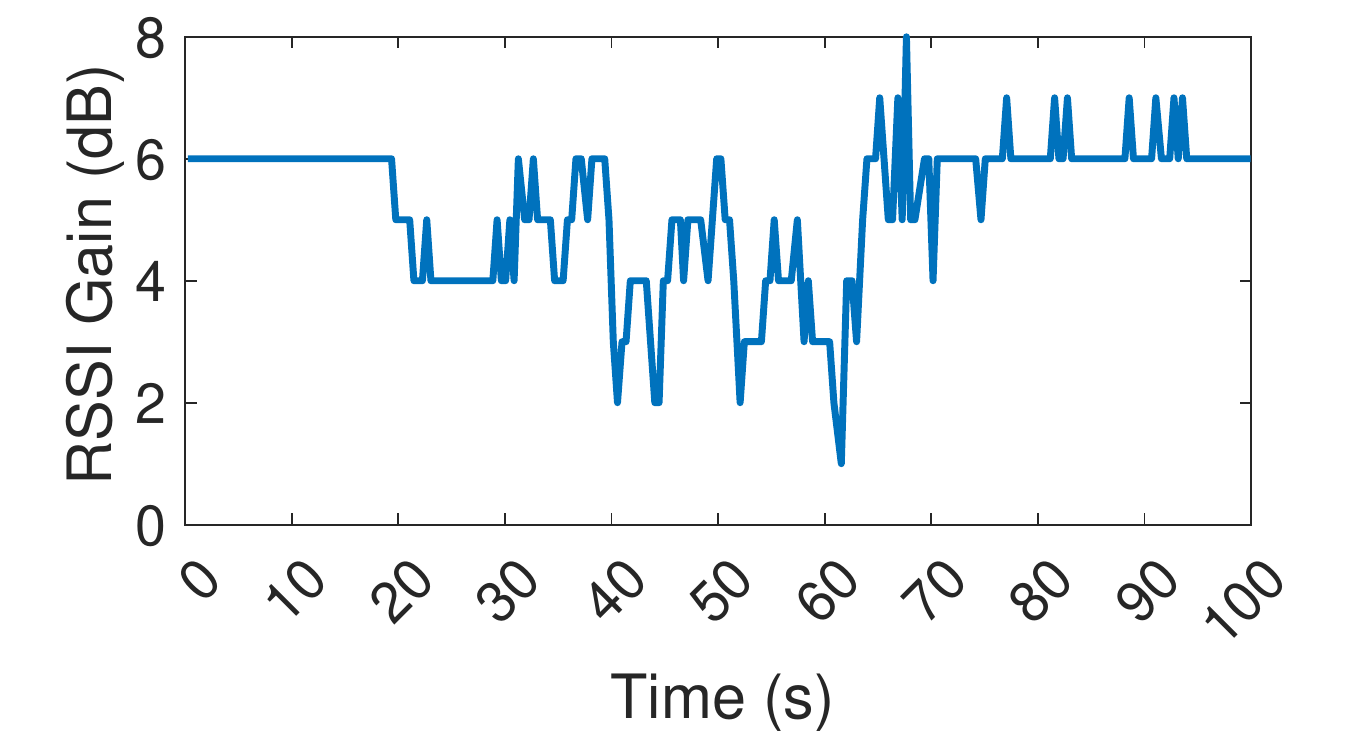}
      \caption{Movement of user.}
      \label{fig:eval-stability-human-walk}
    \end{subfigure}
    \begin{subfigure}{0.48\columnwidth}
      \centering
      \includegraphics[width=\linewidth]{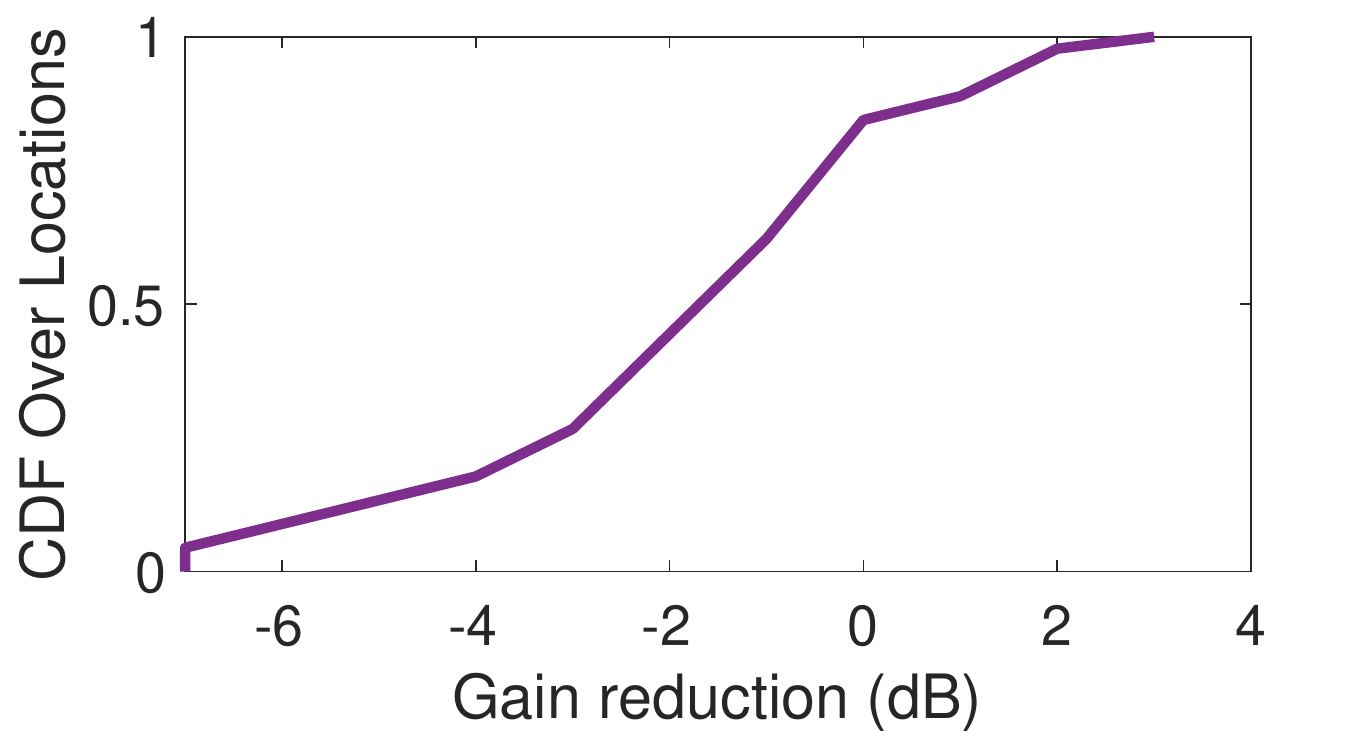}
      \caption{Movement of transmitter.}
      \label{fig:eval-stability-tx-moved}
    \end{subfigure}
    \caption{Performance stability.}
    \label{fig:eval-stability}
  \end{minipage}
\end{figure*}


\subsection{Control Algorithm Microbenchmarks}
\label{sec:eval_micro}

\heading{Convergence speed.}
We now evaluate the time needed to find surface configurations and the speedup relative to brute-force enumeration. \figref{fig:eval-scrolls-opt-time} shows the convergence time for links presented in previous sections. The 
difference between 2.4~GHz and 5~GHz is largely a consequence of the different roll length ranges to be explored and the device-specific feedback reporting speed. Our group sweeping algorithm provides a median of 30\% and up to 3$\times$ speedup 
over enumerating all rolls. Loading a surface configuration cache takes a much shorter time, within a few seconds for the current prototype.

The dominant factor of convergence time is our system's rolling actuation speed. The stepper motors we use provide high rotation precision at low cost, but have a low rotation speed (only around 20~RPM). A better mechanical design would make the system much faster and our design can work with any motor that provides sub-centimeter level precision. To demonstrate this, we assemble a new prototype with another kind of motor\cite{nema17_stepper_motor}. As shown in \figref{fig:eval-scrolls-opt-time}, this new motor achieves around 4$\times$ speedup without any modification to the \name\ control system. We show the time spent to enumerate all rows, which gives an upper bound for convergence time. We believe more engineering efforts can further improve our system's speed, such as using advanced stepper motors to achieve 5$\times$ speedup, and using DC motors for 10$\times$ speedup.


\heading{Performance stability.}
\edit{Lastly, we show \name\ is robust to environmental fluctuation.}
\figref{fig:eval-stability-human-walk} shows the RSSI gain variation over time when a person moves around. 
The person sits near an endpoint for 20~s, and then walks between the endpoints, intentionally blocking the direct path to introduce more variations until 80~s. Our system preserves most of its gain during the period and quickly recovers after the movement stops. We also move objects (such as books) on the table and observe similar results. 
If the endpoint moves only slightly, 
\name\ can still preserve most of its gain without changing its configurations. We set up 15 Wi-Fi links and move the transmitter to 3 nearby locations (around 20~cm away) for each link. 
\figref{fig:eval-stability-tx-moved} shows a median \edit{gain reduction of only 2~dB relative to the original endpoint location}. This is because \name\ mainly relies on \edit{roll}-wise amplitude control, which is less susceptible to channel fluctuations than phase-shift based systems.
\edit{Only when the endpoints move substantially or a large metallic object moves in the vicinity does \name{} need to re-run the control algorithm. }


\section{Related Work}
\label{s:related}

The most relevant work to \name\ concerns reconfigurable antenna/surface designs and network coverage extension leveraging signal-level operations.


\heading{\myedit{Smart surfaces.}} 
\edit{Metasurfaces were initially designed as 2D versions of artificial material with special properties, such as (spatial) negative refraction index~\cite{smith-SRR-2000}, wave front tailoring~\cite{pfeiffer-Huygens-2013}, and temporal reflection~\cite{moussa2023timeReflection}. In recent years, \textit{programmable} metasurfaces or \textit{smart surfaces} have been designed to dynamically reconfigure their signal manipulation behavior. Many proposals~\cite{ris-access2019, visorsurf-wowmom, visorsurf-commsmag2018, guo2020weighted, huang2019reconfigurable, yoo2018enhancing, smartenv-2019, irs-commsmag2020} mainly analyze the theoretical performance of different design approaches. Another}
new line of research has produced early end-to-end prototypes~\cite{laia-nsdi19,press-hotnets17,rfocus,scattermimo,lava-sigcomm21, llama, RFlens,cho2022mmwall} to control signal propagation behavior 
to improve the perceived channel conditions at the receiver endpoints. These focus more on the control plane design \edit{than new hardware primitives}. They do not necessarily involve metasurface hardware, but follow the general approach of a surface-like deployment in the propagation environment. 
LAIA controls the signal phase using phase shifters so that signals add constructively at the receivers; ScatterMIMO reflects the signals to create a virtual AP and boost MIMO throughput; RFocus adopts a simpler design that only performs on-off amplitude control to achieve beamforming effect; LAVA uses active amplify-and-forward elements deployed as a \edit{mesh} on the ceiling and demonstrates standard-agnostic feature. 
Some work explores reflector-like smart surfaces~\cite{nolan2021ros,millimirror}, but lacks programmability to \textit{adapt} to environmental changes.
Recent works also investigate the security implications of metasurfaces~\cite{shaikhanov2022metasurface, li2022protego, chen2022metawave}.
However, none of these designs operates across all sub-6~GHz bands. 
In contrast, \name\ focuses on frequency tunability to provide standard-agnostic, wideband wireless link enhancement, achieving both amplitude and signal phase control with mechanical rolling actuation. 
Further, our system employs a flexible and rollable surface design, following the vision of 
metamorphic surfaces~\cite{metamorphic}.


\heading{Reconfigurable antennas.}
There has been ample work on reconfigurable antennas
~\cite{reconfig-ant-deisgn,reconfig-ant-misc}. Most rely on varactors or PIN diodes for electronic tunability. Such designs only support a limited tunable frequency range, normally \edit{less} than 40\% relative to the center frequency (\secref{sec:design-hardware}), and cannot cover all sub-6GHz bands.
Other designs~\cite{five-band-switch-ant, reconfig-mems-ant} use MEMS switches to switch between multiple frequencies. They are limited to a fixed set of frequency bands and require multiple switches per antenna, which significantly increases the hardware cost, power consumption and antenna size. Alternatively, mechanically reconfigurable antennas~\cite{mechanical-reconfig-ant, reconfigurable-helical-ant,tip-extending-pneumatic-antenna} benefit from the physical movement, and show wideband frequency tunability. However, existing mechanical actuation mechanisms are complex, bulky, and unscalable, especially if incorporated in a smart surface.
\name\ builds on existing work and presents a low-complexity, scalable element design with continuous, wideband frequency tunability for smart surfaces.



\heading{Frequency selective surfaces (FSS).} 
FSS~\cite{munk2003finiteFSS, munk2005FSS} refer to surfaces designed to reflect, transmit or absorb electromagnetic waves based on the frequency, using periodic metallic or dielectric patterns. They work as spatial filters with different frequency domain responses. Recent work also proposes tunable frequency selective surfaces, using varactors~\cite{mias2005varactorFSS,ghosh2017broadbandFSS,ghosh2020dual} or mechanical methods, such as liquid metal~\cite{lei2011FSS-liquid} and an origami surface~\cite{biswas2020origamiFSS}. The tuning mechanisms are similar to what reconfigurable antennas need, and thus face the same issues mentioned above.
More importantly, tunable FSS designs are only tunable \textit{surface}-wise, i.e., they do not provide fine-grained controllability since they are intended to work as filters. Adding fine-grained controllability to them, \edit{if at all possible, would incur overwhelming complexity}. For our purposes, however, we need fine-grained controllability to improve wireless channels with minimal complexity.

\heading{Physical layer relays.}
Physical layer relays (e.g., ~\cite{MoVR, FastForward, DelayForward}) 
can be standard-agnostic in theory. However, in practice, 
existing prototypes tend to rely on extra standard-specific details to operate. They extend the wireless coverage with amplify-and-forward, but for one frequency only and requires substantial circuitry or even a full-duplex radio.

\section{Conclusion} 
\label{s:concl}
As new frequency bands are increasingly adopted in diverse IoT scenarios, continuing the \edit{existing} standard- and frequency-specific coverage provisioning \edit{practice} can incur higher \edit{infrastructure} cost and complexity \edit{than necessary}, for both initial deployment and subsequent upgrades.

This paper presents \name\ as a frequency-tunable smart surface to provide wideband, multi-network coverage enhancement. \name\ comprises a collection of flexible copper strips, whose exposed lengths can be adjusted to tune to the desirable operating frequency. Despite the simple design and low-cost prototypes, \name\ can enhance diverse link setups from LoRa on 900~MHz to Wi-Fi like connectivity on 5~GHz. In particular, its ability to support multiple concurrent links on different \edit{frequency bands} highlights its \edit{potential} 
to amortize cost and complexity across standards and networks.

\begin{acks} 
We thank Leandros Tassiulas for helpful discussion.
We also thank the anonymous reviewers and our shepherd for their insightful comments and suggestions.
This work is partially funded by the National Science Foundation under Grant No. 2112562.
\end{acks}

\balance 
\bibliographystyle{concise2}
\bibliography{biblio}

\begin{thebibliography}{10}
\expandafter\ifx\csname urlstyle\endcsname\relax
  \providecommand{\doi}[1]{doi:\discretionary{}{}{}#1}\else
  \providecommand{\doi}{doi:\discretionary{}{}{}\begingroup
  \urlstyle{rm}\Url}\fi

\bibitem{stepper_motor}
{28BYJ-48 Stepper Motor}.
\newblock
  \url{https://www.mouser.com/datasheet/2/758/stepd-01-data-sheet-1143075.pdf}.

\bibitem{c-band-att-verizon}
5g c-band: Verizon and at\&t are flipping on the switch in the us.
\newblock
  \url{https://www.cnn.com/2022/01/19/tech/c-band-5g-att-verizon-rollout/index.html},.

\bibitem{hfss}
{Ansys HFSS 3D High Frequency Simulation Software}.
\newblock \url{https://www.ansys.com/products/electronics/ansys-hfss}.

\bibitem{empower-tech-roadmap}
{EMPOWER: Final technology roadmap for advanced wireless}.
\newblock
  \url{https://www.advancedwireless.eu/wp-content/uploads/2022/02/Del2-5-Final-technology-roadmap.pdf},.

\bibitem{esp32-devkitc}
{Epressif ESP32-DevKitC Board}.
\newblock \url{https://www.espressif.com/en/products/devkits/esp32-devkitc}.

\bibitem{esp-32s}
{ESP-WROOM-32 ESP-32S Dev Board}.
\newblock \url{http://www.hiletgo.com/ProductDetail/1906566.html}.

\bibitem{usrp-n210}
Ettus research usrp n210.
\newblock \url{https://www.ettus.com/all-products/un210-kit/},.

\bibitem{vert2450}
{Ettus Research VERT2450 Antenna}.
\newblock \url{https://www.ettus.com/all-products/vert2450/}.

\bibitem{5g-c-band}
{FCC Expands Flexible Use of the C-band for 5G}.
\newblock
  \url{https://www.fcc.gov/document/fcc-expands-flexible-use-c-band-5g}.

\bibitem{tv-white-space}
{FCC TV white space}.
\newblock \url{https://www.fcc.gov/general/white-space}.

\bibitem{esp32-lora}
{Heltec ESP32 WiFi LoRa Board}.
\newblock \url{https://heltec.org/project/wifi-lora-32/}.

\bibitem{flex600_antenna}
{LairdConnect Revie Flex Series Cellular Antennas}.
\newblock
  \url{https://www.lairdconnect.com/documentation/datasheet-revie-flex-600}.

\bibitem{nema17_stepper_motor}
{NEMA-17 Stepper Motor}.
\newblock \url{https://www.adafruit.com/product/324}.

\bibitem{next-g-alliance-report}
{Next G Alliance Report: Roadmap to 6G}.
\newblock
  \url{https://nextgalliance.org/wp-content/uploads/2022/02/NextGA-Roadmap.pdf},.

\bibitem{varactor-spec}
Smv123x series varactor.
\newblock \url{https://www.skyworksinc.com/Products/Diodes/SMV1232-Series},.

\bibitem{shift-reg}
{SN74HC595 8-Bit Shift Register}.
\newblock \url{https://www.ti.com/lit/ds/symlink/sn74hc595.pdf}.

\bibitem{motor-lifespan}
Step motor guide.
\newblock
  \url{https://www.anaheimautomation.com/manuals/forms/stepper-motor-guide.php},.

\bibitem{motor_drive_chip}
{ULN2003A transistor array chip}.
\newblock \url{https://www.ti.com/product/ULN2003A}.

\bibitem{what-is-c-band}
What is c-band, and what does it mean for the future of 5g?
\newblock \url{https://www.pcmag.com/news/what-is-c-band},.

\bibitem{MoVR}
O.~Abari, D.~Bharadia, A.~Duffield, D.~Katabi.
\newblock {Enabling High-quality Untethered Virtual Reality}.
\newblock \textit{Proceedings of USENIX NSDI}, 2017.

\bibitem{multiband-antennas}
H.~F. Abutarboush.
\newblock \textit{Fixed and reconfigurable multiband antennas}.
\newblock Ph.D. thesis, 2011.

\bibitem{rfocus}
V.~Arun, H.~Balakrishnan.
\newblock {RFocus: Practical Beamforming for Small Devices}.
\newblock \textit{Symposium on Networked Systems Design and Implementation
  (NSDI)}, 1047--1061. USENIX, 2020.

\bibitem{ris-access2019}
E.~{Basar}, M.~{Di Renzo}, J.~{De Rosny}, M.~{Debbah}, M.~{Alouini},
  R.~{Zhang}.
\newblock {Wireless Communications Through Reconfigurable Intelligent
  Surfaces}.
\newblock \textit{IEEE Access}, \textbf{7}, 2019.

\bibitem{FastForward}
D.~Bharadia, S.~Katti.
\newblock {FastForward: Fast and Constructive Full Duplex Relays}.
\newblock \textit{Proceedings of ACM SIGCOMM}, 2014.

\bibitem{frequency-agile-microstrip}
P.~Bhartia, I.~Bahl.
\newblock A frequency agile microstrip antenna.
\newblock \textit{1982 Antennas and Propagation Society International
  Symposium}, vol.~20, 304--307. IEEE, 1982.

\bibitem{biswas2020origamiFSS}
A.~Biswas, C.~L. Zekios, S.~V. Georgakopoulos.
\newblock Transforming single-band static fss to dual-band dynamic fss using
  origami.
\newblock \textit{Scientific Reports}, \textbf{10}(1), 1--12, 2020.

\bibitem{tip-extending-pneumatic-antenna}
L.~H. Blumenschein, L.~T. Gan, J.~A. Fan, A.~M. Okamura, E.~W. Hawkes.
\newblock A tip-extending soft robot enables reconfigurable and deployable
  antennas.
\newblock \textit{IEEE Robotics and Automation Letters}, \textbf{3}(2),
  949--956, 2018.

\bibitem{five-band-switch-ant}
K.~R. Boyle, P.~G. Steeneken.
\newblock A five-band reconfigurable pifa for mobile phones.
\newblock \textit{IEEE Transactions on Antennas and Propagation},
  \textbf{55}(11), 3300--3309, 2007.

\bibitem{llama}
L.~Chen, W.~Hu, K.~Jamieson, X.~Chen, D.~Fang, J.~Gummeson.
\newblock {Pushing the physical limits of IoT devices with programmable
  metasurfaces}.
\newblock \textit{Symposium on Networked Systems Design and Implementation
  (NSDI)}, 425--438. USENIX, 2021.

\bibitem{chen2022metawave}
X.~Chen, Z.~Li, B.~Chen, Y.~Zhu, C.~X. Lu, Z.~Peng, F.~Lin, W.~Xu, K.~Ren,
  C.~Qiao.
\newblock Metawave: Attacking mmwave sensing with meta-material-enhanced tags.
\newblock \textit{The 30th Network and Distributed System Security (NDSS)
  Symposium}. The Internet Society, 2023.

\bibitem{cho2022mmwall}
K.~W. Cho, M.~H. Mazaheri, J.~Gummeson, O.~Abari, K.~Jamieson.
\newblock mmwall: A transflective metamaterial surface for mmwave networks.
\newblock \textit{arXiv preprint arXiv:2209.11554}, 2022.

\bibitem{reconfig-ant-deisgn}
J.~Costantine, Y.~Tawk, S.~E. Barbin, C.~G. Christodoulou.
\newblock Reconfigurable antennas: Design and applications.
\newblock \textit{Proceedings of the IEEE}, \textbf{103}(3), 424--437, 2015.
\newblock \doi{10.1109/JPROC.2015.2396000}.

\bibitem{reconfigurable-helical-ant}
J.~Costantine, Y.~Tawk, C.~Christodoulou.
\newblock Reconfigurable deployable antennas for space communications.
\newblock \textit{2014 International Workshop on Antenna Technology: Small
  Antennas, Novel EM Structures and Materials, and Applications (iWAT)},
  151--154. IEEE, 2014.

\bibitem{mechanical-reconfig-ant}
J.~Costantine, Y.~Tawk, J.~Woodland, N.~Flaum, C.~G. Christodoulou.
\newblock Reconfigurable antenna system with a movable ground plane for
  cognitive radio.
\newblock \textit{IET Microwaves, Antennas \& Propagation}, \textbf{8}(11),
  858--863, 2014.

\bibitem{scattermimo}
M.~Dunna, C.~Zhang, D.~Sievenpiper, D.~Bharadia.
\newblock {ScatterMIMO: enabling virtual MIMO with smart surfaces}.
\newblock \textit{Proceedings of ACM Mobicom}, 1--14. ACM, 2020.

\bibitem{RFlens}
C.~Feng, X.~Li, Y.~Zhang, X.~Wang, L.~Chang, F.~Wang, X.~Zhang, X.~Chen.
\newblock Rflens: Metasurface-enabled beamforming for iot communication and
  sensing.
\newblock \textit{Proceedings of ACM Mobicom}, 2021.

\bibitem{ghosh2017broadbandFSS}
S.~Ghosh, K.~V. Srivastava.
\newblock Broadband polarization-insensitive tunable frequency selective
  surface for wideband shielding.
\newblock \textit{IEEE Transactions on Electromagnetic Compatibility},
  \textbf{60}(1), 166--172, 2017.

\bibitem{ghosh2020dual}
S.~Ghosh, K.~V. Srivastava.
\newblock A dual-band tunable frequency selective surface with independent
  wideband tuning.
\newblock \textit{IEEE Antennas and Wireless Propagation Letters},
  \textbf{19}(10), 1808--1812, 2020.

\bibitem{guo2020weighted}
H.~Guo, Y.-C. Liang, J.~Chen, E.~G. Larsson.
\newblock Weighted sum-rate maximization for reconfigurable intelligent surface
  aided wireless networks.
\newblock \textit{IEEE Transactions on Wireless Communications},
  \textbf{19}(5), 3064--3076, 2020.

\bibitem{effectivesnr-sigcomm10}
D.~Halperin, W.~Hu, A.~Sheth, D.~Wetherall.
\newblock Predictable 802.11 packet delivery from wireless channel
  measurements.
\newblock \textit{ACM SIGCOMM Computer Communication Review}, \textbf{40}(4),
  159--170, 2010.

\bibitem{reconfig-ant-misc}
R.~L. Haupt, M.~Lanagan.
\newblock Reconfigurable antennas.
\newblock \textit{IEEE Antennas and Propagation Magazine}, \textbf{55}(1),
  49--61, 2013.

\bibitem{DelayForward}
K.-C. Hsu, K.~C.-J. Lin, H.-Y. Wei.
\newblock {Full-duplex Delay-and-forward Relaying}.
\newblock \textit{Proceedings of ACM MobiHoc}, 2016.

\bibitem{beyond-massive-mimo}
S.~Hu, F.~Rusek, O.~Edfors.
\newblock Beyond massive mimo: The potential of data transmission with large
  intelligent surfaces.
\newblock \textit{IEEE Transactions on Signal Processing}, \textbf{66}(10),
  2746--2758, 2018.

\bibitem{huang2019reconfigurable}
C.~Huang, A.~Zappone, G.~C. Alexandropoulos, M.~Debbah, C.~Yuen.
\newblock {Reconfigurable intelligent surfaces for energy efficiency in
  wireless communication}.
\newblock \textit{IEEE Transactions on Wireless Communications},
  \textbf{18}(8), 4157--4170, 2019.

\bibitem{lei2011FSS-liquid}
B.~J. Lei, A.~Zamora, T.~F. Chun, A.~T. Ohta, W.~A. Shiroma.
\newblock A wideband, pressure-driven, liquid-tunable frequency selective
  surface.
\newblock \textit{IEEE microwave and wireless components letters},
  \textbf{21}(9), 465--467, 2011.

\bibitem{li2022protego}
X.~Li, C.~Feng, F.~Song, C.~Jiang, Y.~Zhang, K.~Li, X.~Zhang, X.~Chen.
\newblock Protego: securing wireless communication via programmable
  metasurface.
\newblock \textit{Proceedings of ACM MobiCom}, 55--68, 2022.

\bibitem{laia-nsdi19}
Z.~Li, Y.~Xie, L.~Shangguan, R.~I. Zelaya, J.~Gummeson, W.~Hu, K.~Jamieson.
\newblock {Towards Programming the Radio Environment with Large Arrays of
  Inexpensive Antennas}.
\newblock \textit{Proceedings of USENIX NSDI}, 285--300. USENIX, 2019.

\bibitem{visorsurf-commsmag2018}
C.~{Liaskos}, S.~{Nie}, A.~{Tsioliaridou}, A.~{Pitsillides}, S.~{Ioannidis},
  I.~{Akyildiz}.
\newblock {A New Wireless Communication Paradigm through Software-Controlled
  Metasurfaces}.
\newblock \textit{IEEE Communications Magazine}, \textbf{56}(9), 2018.

\bibitem{visorsurf-wowmom}
C.~Liaskos, S.~Nie, A.~Tsioliaridou, A.~Pitsillides, S.~Ioannidis, I.~Akyildiz.
\newblock {Realizing wireless communication through software-defined
  hypersurface environments}.
\newblock \textit{Proceedings of IEEE WoWMoM}, 2018.

\bibitem{highly-stretchable-circuits}
S.~Liu, D.~S. Shah, R.~Kramer-Bottiglio.
\newblock Highly stretchable multilayer electronic circuits using biphasic
  gallium-indium.
\newblock \textit{Nature Materials}, \textbf{20}(6), 851--858, 2021.

\bibitem{mias2005varactorFSS}
C.~Mias.
\newblock Varactor-tunable frequency selective surface with
  resistive-lumped-element biasing grids.
\newblock \textit{IEEE microwave and wireless components letters},
  \textbf{15}(9), 570--572, 2005.

\bibitem{moussa2023timeReflection}
H.~Moussa, G.~Xu, S.~Yin, E.~Galiffi, Y.~Ra’di, A.~Al{\`u}.
\newblock Observation of temporal reflection and broadband frequency
  translation at photonic time interfaces.
\newblock \textit{Nature Physics}, 1--6, 2023.

\bibitem{munk2003finiteFSS}
B.~A. Munk.
\newblock \textit{Finite antenna arrays and FSS}.
\newblock John Wiley \& Sons, 2003.

\bibitem{munk2005FSS}
B.~A. Munk.
\newblock \textit{Frequency selective surfaces: theory and design}.
\newblock John Wiley \& Sons, 2005.

\bibitem{nolan2021ros}
J.~Nolan, K.~Qian, X.~Zhang.
\newblock {RoS: Passive Smart Surface for Roadside-to-Vehicle Communication}.
\newblock \textit{Proceedings of ACM SIGCOMM}, 165--178. ACM, 2021.

\bibitem{pfeiffer-Huygens-2013}
C.~Pfeiffer, A.~Grbic.
\newblock Metamaterial {{Huygens}}' {{Surfaces}}: {{Tailoring Wave Fronts}}
  with {{Reflectionless Sheets}}.
\newblock \textit{Physical Review Letters}, \textbf{110}(19), 197,401, 2013.
\newblock \doi{10.1103/PhysRevLett.110.197401}.

\bibitem{millimirror}
K.~Qian, X.~Zhang.
\newblock Millimirror: 3d printed reflecting surface for millimeter-wave
  coverage expansion.
\newblock \textit{Proceedings of ACM MobiCom}, 2022.

\bibitem{smartenv-2019}
M.~D. Renzo, M.~Debbah, D.-T. Phan-Huy, A.~Zappone, M.-S. Alouini, C.~Yuen,
  V.~Sciancalepore, G.~C. Alexandropoulos, J.~Hoydis, H.~Gacanin, J.~de~Rosny,
  A.~Bounceur, G.~Lerosey, M.~Fink.
\newblock {Smart radio environments empowered by reconfigurable AI
  meta-surfaces: An idea whose time has come}.
\newblock \textit{EURASIP Journal on Wireless Communications and Networking
  volume}, 2019.

\bibitem{shaikhanov2022metasurface}
Z.~Shaikhanov, F.~Hassan, H.~Guerboukha, D.~Mittleman, E.~Knightly.
\newblock Metasurface-in-the-middle attack: from theory to experiment.
\newblock \textit{Proceedings of ACM WiSec}, 257--267, 2022.

\bibitem{freeSpectrum-Biden-2023}
D.~Shepardson.
\newblock Biden administration looks to free up wireless spectrum for advanced
  technology.
\newblock \textit{Reuters}, 2023.

\bibitem{smith-SRR-2000}
D.~R. Smith, W.~J. Padilla, D.~C. Vier, S.~C. {Nemat-Nasser}, S.~Schultz.
\newblock Composite {{Medium}} with {{Simultaneously Negative Permeability}}
  and {{Permittivity}}.
\newblock \textit{Physical Review Letters}, \textbf{84}(18), 4184--4187, 2000.
\newblock \doi{10.1103/PhysRevLett.84.4184}.

\bibitem{antenna-theory-design}
W.~L. Stutzman, G.~A. Thiele.
\newblock \textit{Antenna theory and design}.
\newblock John Wiley \& Sons, 2012.

\bibitem{6g-vision-hexa}
M.~A. Uusitalo, P.~Rugeland, M.~R. Boldi, E.~C. Strinati, P.~Demestichas,
  M.~Ericson, G.~P. Fettweis, M.~C. Filippou, A.~Gati, M.-H. Hamon,
  \textit{et~al.}
\newblock 6g vision, value, use cases and technologies from european 6g
  flagship project hexa-x.
\newblock \textit{IEEE Access}, \textbf{9}, 160,004--160,020, 2021.

\bibitem{press-hotnets17}
A.~Welkie, L.~Shangguan, J.~Gummeson, W.~Hu, K.~Jamieson.
\newblock {Programmable Radio Environments for Smart Spaces}.
\newblock \textit{Proceedings of ACM HotNets}. ACM, 2017.

\bibitem{irs-commsmag2020}
Q.~{Wu}, R.~{Zhang}.
\newblock {Towards Smart and Reconfigurable Environment: Intelligent Reflecting
  Surface Aided Wireless Network}.
\newblock \textit{IEEE Communications Magazine}, \textbf{58}(1), 2020.

\bibitem{reconfig-mems-ant}
T.~Yamagajo, Y.~Koga.
\newblock Frequency reconfigurable antenna with mems switches for mobile
  terminals.
\newblock \textit{2011 IEEE-APS Topical Conference on Antennas and Propagation
  in Wireless Communications}, 1213--1216. IEEE, 2011.

\bibitem{yoo2018enhancing}
I.~Yoo, M.~F. Imani, T.~Sleasman, H.~D. Pfister, D.~R. Smith.
\newblock {Enhancing capacity of spatial multiplexing systems using
  reconfigurable cavity-backed metasurface antennas in clustered MIMO
  channels}.
\newblock \textit{IEEE Transactions on Communications}, \textbf{67}(2),
  1070--1084, 2018.

\bibitem{metamorphic}
R.~I. Zelaya, R.~Ma, W.~Hu.
\newblock Towards 6g and beyond: Smarten everything with metamorphic surfaces.
\newblock \textit{Proceedings of ACM HotNets}, 155--162, 2021.

\bibitem{lava-sigcomm21}
R.~I. Zelaya, W.~Sussman, J.~Gummeson, K.~Jamieson, W.~Hu.
\newblock {LAVA: Fine-Grained 3D Indoor Wireless Coverage for Small IoT
  Devices}.
\newblock \textit{Proceedings of ACM SIGCOMM}, 123--136. ACM, 2021.

\end{thebibliography}

\end{document}